\documentclass[3pt,twoside]{amsart}
\usepackage{amssymb,mathrsfs}
\usepackage{tabularx}
\usepackage{epsfig}
\usepackage[hypertex,linkcolor=red]{hyperref}
\def\<{\langle}\def\>{\rangle}\def\Reals{\mathbb R}\def\Cmplx{\mathbb C}\def\Field{\mathbb F}
\def\set#1{{\sf #1}}
\def\transp#1{{#1}^{\, t}}
\def\System{\set{S}}\def\Stset{{\mathfrak S}}\def\Trnset{{\mathfrak T}}\def\Cntset{{\mathfrak E}} 
\def\Span{\set{Span}}\def\Bnd#1{\set{Lin(#1)}}\def\Bndp#1{\set{Lin}_+(#1)}\def\Erays{\set{Erays}}
\def\sH{\set{H}}\def\qed{$\,\blacksquare$\par}\def\n#1{|\!|#1|\!|}
\def\eff{{\rm eff}}\renewcommand{\geq}{\geqslant}\renewcommand{\leq}{\leqslant}
\def\Postulate#1#2#3{\bigskip\par\noindent{\bf Postulate #1 (#2).$\;$}{\em #3}\bigskip\par}
\def\MPostulate#1#2#3{\bigskip\par\noindent{\bf Mathematical Postulate #1 (#2).$\;$}{\em #3}\bigskip\par}
\newtheorem{lemma}{Lemma}\newtheorem{theorem}{Theorem}\newtheorem{corollary}{Corollary}
\def\Proof{\medskip\par\noindent{\bf Proof. }}
\def\trnsfrm#1{\mathscr #1}
\def\tA{\trnsfrm A}\def\tB{\trnsfrm B}\def\tC{\trnsfrm C}\def\tD{\trnsfrm D}
\def\tI{\trnsfrm I}\def\tT{\trnsfrm T}\def\tS{\trnsfrm S}\def\tZ{\trnsfrm Z}\def\tU{\trnsfrm U}

\def\AA{\mathbb A}\def\AB{\mathbb B}\def\AC{\mathbb C}\def\AD{\mathbb D}
\def\AM{\mathbb M}\def\AL{\mathbb L}\def\AS{\mathbb S}
\def\Stset{{\mathfrak S}}\def\alg#1{{\mathcal #1}}\def\map#1{{\mathcal #1}}
\def\Tr{\operatorname{Tr}}

\begin{document}
\title[Probabilistic theories: what is special about Quantum Mechanics?]{Probabilistic theories:\\
  what is special about Quantum Mechanics?}\author{Giacomo Mauro D'Ariano} \email{dariano@unipv.it}
\address{{\em QUIT} Group, http://www.qubit.it, Istituto Nazionale di Fisica Nucleare, Sezione di
  Pavia\\ Dipartimento di Fisica ``A. Volta'', via Bassi 6, I-27100 Pavia, Italy, and \\ Department
  of Electrical and Computer Engineering, Northwestern University, Evanston, IL 60208}
\begin{abstract} 
  Quantum Mechanics (QM) is a very special probabilistic theory, yet we don't know which operational
  principles make it so. All axiomatization attempts suffer at least one postulate of a mathematical
  nature.  Here I will analyze the possibility of deriving QM as the mathematical representation of a
  {\em fair operational framework}, {\it i.e.} a set of rules which allows the experimenter to make
  predictions on future {\em events} on the basis of suitable {\em tests}, {\it e.g.} without interference
  from uncontrollable sources. Two postulates need to be satisfied by any fair operational
  framework: NSF: {\em no-signaling from the future}---for the possibility of making predictions on
  the basis of past tests; PFAITH: {\em existence of a preparationally faithful state}---for the
  possibility of preparing any state and calibrating any test. I will show that all theories
  satisfying NSF admit a C${}^*$-algebra representation of events as linear transformations of
  effects. Based on a very general notion of dynamical independence, it is easy to see that all such
  probabilistic theories are {\em non-signaling without interaction} ({\em non-signaling} for
  short)---another requirement for a fair operational framework.  Postulate PFAITH then implies the
  {\em local observability principle}, the tensor-product structure for the linear spaces of states
  and effects, the impossibility of bit commitment and additional features, such an operational
  definition of transpose, a scalar product for effects, weak-selfduality of the theory, and more.
  Dual to Postulate PFAITH an analogous postulate for effects would give additional quantum
  features, such as teleportation.  However, all possible consequences of these postulates still
  need to be investigated, and it is not clear yet if we can derive QM from the present postulates
  only.

  What is special about QM is that also {\em effects make a C${}^*$-algebra}. More precisely, this
  is true for all hybrid quantum-classical theories, corresponding to QM plus super-selection rules.
  However, whereas the sum of effects can be operationally defined, the notion of effect abhors any
  kind of composition. Based on the natural postulate of atomicity of evolution (AE) one can define
  composition of effects by identifying them with atomic events through the Choi-Jamiolkowski
  isomorphism (CJ). In this way the quantum-classical hybrid is selected within the large arena of
  non-signaling probabilistic theories, including the {\em Popescu-Rohrlich boxes}. The CJ
  isomorphism looks natural in an operational context, and it is hoped that it will soon be
  converted into an operational principle.
\end{abstract}
\footnote{To appear in {\em Philosophy of Quantum Information and Entanglement}, Eds A. Bokulich and G. Jaeger (Cambridge University Press, Cambridge UK)}   
\date{version of \today}
\maketitle
\begin{itemize}
\item[] \hfill {\em To my friend and mentor, Professor Attilio Rigamonti.}
\end{itemize}
\vskip 1cm
\bigskip
\begin{itemize}
\item[] {\em Unperformed experiments have no results.}
\item[] \hfill ---{\bf Asher Peres}
\end{itemize}
\newpage
\section{Introduction}
After more than a century from its birth, Quantum Mechanics (QM) remains mysterious. We still don't
have general principles from which to derive its remarkable mathematical framework, as it happened
for the amazing Lorentz's transformations, lately re-derived by Einstein's from invariance of
physical laws on inertial frames and from constancy of the speed of light. 

Despite the utmost relevance of the problem of deriving QM from operational principles, research
efforts in this direction have been sporadic. The deepest of the early attacks on the problem were
the works of Birkoff, von Neumann, Jordan, and Wigner, attempting to derive QM from a set of axioms
with as much physical significance as possible \cite{Birkhoff:1936p6,JordanP:1934p29}.  The general
idea in Ref.  \cite{Birkhoff:1936p6} is to regard QM as a new kind of {\em prepositional calculus},
a proposal that spawned the research line of {\em quantum logic}, based on von Neumann's observation
that the two-valued observables---represented in his formulation of QM by orthogonal projection
operators---constitute a kind of ``logic'' of experimental propositions. After a hiatus of two
decades of neglect, interest in quantum logic was revived by Varadarajan
\cite{VaradarajanV1962p217}, and most notably by Mackey \cite{MackeyG1963}, who axiomatized QM
within an operational framework, with the single exception of an admittedly {\em ad hoc} postulate,
which represents the propositional-calculus mathematically in form of an orthomodular lattice. The
most significant extension of Mackey's work is the general representation theorem of Piron
\cite{PironC1976}.

In the early work \cite{JordanP:1934p29}, Jordan, von Neumann, and Wigner considered the possibility
of a commutative algebra of observables, with a product which needs only to define squares and sums
of observables---the so-called {\em Jordan product} of observables $a$ and $b$: $a\circ
b:=(a+b)^2-a^2-b^2$. However, such a product is generally non-associative and non-distributive with
respect to the sum, and the quantum formalism follows only with additional axioms with no clear
physical significance---{\it e.g.} a distributivity axiom for the Jordan product. Segal
\cite{Segal:1947p1408} later constructed an (almost) fully operational framework (with no
experimental definition of the sum of observables) which allows generally non-distributive algebras
of observables, but with a resulting mathematical framework largely more general than QM.  As a
result of this line of investigation, the purely algebraic formulation of QM gathered popularity versus the
original Hilbert-space axiomatization.

In the algebraic axiomatization of QM, a physical system is defined by its $C^*$-algebra of
observables (with identity), and the states of the system are identified with normalized positive
linear functionals over the algebra, corresponding to the probability rules of measurements of
observables.  Indeed, the $C^*$-algebra of observables is more general than QM, since it includes
Classical Mechanics as a special case, and generally describes any quantum-classical hybrid,
equivalent to QM with super-selection rules. Since in practice two observables are not
distinguishable if they always exhibit the same probability distributions, at the operational level
one can always take the set of states as {\em observable-separating}---in the sense that there are
no different observables having the same probability distribution for all states. Conversely the set
of observables is {\em state-separating}, {\it i.e.} there are no different states corresponding to the same
probability distribution for all observables. Notice that, in principle, there exist different
observables with the same expectation for all states, but higher moments will be 
different.\footnote{This is not the case when one considers only {\em sharp observables}, for which
  there always exists a state such that the expectation of any function of the observable equals the
  function of the expectation. However, operationally we cannot rely on such concept to define the
  general notion of observable, since we cannot reasonably assume its feasibility (actual
  measurements are non-sharp).\label{n:sharp}}

The algebra of observables is generally considered to be more ``operational'' than the usual
Hilbert-space axiomatization, however, little is gained more than a representation-independent
mathematical framework.  Indeed, the algebraic framework is unable to provide operational rules for
how to measure sums and products of non-commuting observables.\footnote{The spectrum of the sum is
  generally different from the sum of the spectra of the addenda, {\it e.g.} the spectra of $xp_y$ and
  $yp_x$ are both $\Reals$, whereas the angular momentum component $xp_y-yp_x$ has discrete
  spectrum. The same is true for the product.}  The sum of two observables cannot be given an
operational meaning, since a procedure involving the measurements of the two addenda would
unavoidably assume that their respective measurements are jointly executable on the same
system---{\it i.e.} the observables are {\em compatible}.  The same reasoning holds for the product of two
observables. A sum-observable defined as the one having expectation equal to the sum of expectations
for all states \cite{StrocchiF2005} is clearly not unique, due to the existence of observables
having the same expectation for all states, but with different higher moments. The only well defined
procedures are those involving single observables, such as the measurement of a {\em function of a
  single observable}, which operationally consists in just taking the function of the outcome.

The Jordan symmetric product has been regarded as a great advance in view of an operational
axiomatization, since, in addition to being Hermitian (observables are Hermitian), is defined only
in terms of squares and sums of observables---{\it i.e.} without products. The definition of $a\circ b$,
however, still relies on the notion of sum of observables, which has no operational meaning.
Remarkably, Alfsen and Shultz \cite{AlfsenE2001,AlfsenE2002} succeeded in deriving the Jordan
product from solely geometrical properties of the convex set of states---{\it e.g.} orientability and faces
shaped as Euclidean balls---however, again with no operational meaning. The problem with the Jordan
product is that, in addition to not necessarily being associative, it is not even distributive, as
the reader can easily check himself. It turns out that, modulo a few topological assumptions, the
Jordan algebras can be embedded in the algebra $\Bnd{H}$ of operators over the Hilbert space
$\sH$, whereby $a\circ b=ab+ba$. Such assumptions, however, are still not operational. For further
critical overview of these earlier attempts to an operational axiomatization of QM, the reader is
also directed to the recent books of Strocchi \cite{StrocchiF2005} and Thirring
\cite{ThirringW2004}.

After a long suspension of research on the axiomatic approach---notably interrupted by the work of Ludwig and
his school \cite{Ludwig-axI}---in the last years the new field of Quantum Information has renewed
the interest on the problem of operational axiomatization of QM, boosted by the new experience on
multipartite systems and {\em entanglement}. In his seminal paper \cite{Hardy:2001p595} Hardy
derived QM from five ``reasonable axioms'', which, more than being truly operational, are motivated
on the basis of simplicity and continuity criteria, with the assumption of a finite number of
perfectly discriminable states. His axiom 4, however, is still purely mathematical, and is directly
related to the tensor product rule for composite systems. In another popular paper
\cite{Clifton:2003p91}, Clifton, Bub, and Halvorson have shown how three fundamental
information-theoretic constraints---(a) the no-signaling, (b) the no-broadcasting, (c) the
impossibility of unconditionally secure bit commitment---suffice to entail that the observables and
state space of a physical theory are quantum-mechanical.  Unfortunately, the authors already started
from a C${}^*$-algebraic framework for observables, which, as already discussed, has little
operational basis, and already coincides with the quantum-classical hybrid.  Therefore, more than
deriving QM, their informational principles just force the C${}^*$-algebra of observables to be
non-Abelian.

In Ref.  \cite{darianoVax2006}\footnote{Most of the results of Ref.  \cite{darianoVax2006} were
  originally conjectured in Refs.  \cite{darianoVax2005} and \cite{dariano-losini2005}.} I showed
how it is possible to derive the formulation of QM in terms of observables represented as Hermitian
operators over Hilbert spaces with the right dimensions for the tensor product, starting from few
operational axioms. However, it is not clear yet if such framework is sufficient to identify QM (or
the quantum-classical hybrid) as the only probabilistic theory resulting from axioms. Later, in Refs.
\cite{darianoVax2007,QCM07,beyondthequantum} I have shown how a C${}^*$-algebraic framework for
transformations (not for observables!)  naturally follows from an operational probabilistic
framework.

A very recent and promising direction for attacking the problem of QM axiomatization consists in
positioning QM within the landscape of general probabilistic theories, including theories with
non-local correlations stronger than the quantum ones, {\it e.g.} for the Popescu-Rohrlich boxes (PR-boxes)
\cite{Popescu:1994p1581}. Such theories have correlations that are ``stronger'' than the quantum
ones---in the sense that they violate the quantum Cirel'son bound
\cite{Cirelson:1980p1582}---although they are still non-signaling, thus revealing the fortuitousness
of the peaceful coexistence of QM and Special Relativity, in contrast with the claimed implication
of QM linearity from the no-signaling condition \cite{Simon:2001p199}. Within the framework of the
PR-boxes general versions of the no-cloning and no-broadcasting theorems have been proved
\cite{Barnum:2007p179}. In Ref.  \cite{Barrett:2005p3251} it has been shown that certain features
generally thought of as specifically quantum, are indeed present in all except classical theories.
These include the non-unique decomposition of a mixed state into pure states, disturbance on
measurement (related to the possibility of secure key distribution), and no-cloning. More recently,
necessary and sufficient conditions have been established for teleportation \cite{Barnum:2008p1588},
{\it i.e.} for reconstructing the state of a system on a remote identical system, using only local
operations and joint states. In all these works Quantum Information has inspired task-oriented
axioms to be considered in a general operational framework that can incorporate QM, classical
theory, and other non-signaling probabilistic theories (for an illustration of this general point of
view see also Ref.  \cite{Barnum:2006p606}).

In the present paper I will consider the possibility of deriving QM as the mathematical
representation of a {\em fair operational framework}, {\it i.e.} a set of rules which allows the
experimenter to make predictions regarding future {\em events} on the basis of suitable {\em tests}, in a
spirit close to Ludwig's axiomatization \cite{Ludwig-axI}. {\em States} are simply the compendia of
probabilities for all possible outcomes of any test. I will consider a very general class of
probabilistic theories, and examine the consequences of two Postulates that need to be satisfied by
any fair operational framework: \medskip
\begin{itemize}
\item[NSF:] {\em no-signaling from the future}, implying that it is possible to make
    predictions based on present tests;
  \item[PFAITH:] {\em existence of preparationally faithful states}, implying the possibility of
    preparing any state and calibrating any test.
\end{itemize}
\medskip

NSF is implicit in the definition itself of conditional probabilities for cascade-tests, entailing
that {\em events are identified with transformations}, whence {\em evolution is identified with
  conditioning}.  As we will see, such identifications lead to the notion of {\em effect} of Ludwig,
{\it i.e.} the equivalence class of events occurring with the same probability for all states. I will show
how we can introduce operationally a linear-space structure for effects. I will then show how all
theories satisfying NSF admit a C${}^*$-algebra representation of events as linear transformations
of effects.

Based on a very general notion of dynamical independence, entailing the definition of {\em marginal
  state}, it is immediately seen that all these theories are {\em non-signaling}, which is the
current way of saying that the theories satisfy the principle of {\em Einstein locality}, namely
that there can be no detectable effect on a system due to anything done to another non-interacting
system. This is clearly another requirement for a fair operational framework. Postulate PFAITH then
implies the {\em local observability principle}, namely the possibility of achieving an
informationally complete test using only local tests---another requirement for a fair operational
framework. The same postulate also implies many other features that are typically quantum, such as
the tensor-product structure for the linear spaces of states and effects, the isomorphism of cones
of states and effects---a weaker version of the quantum selfduality---impossibility of bit
commitment (which in Ref. \cite{Clifton:2003p91} we remind it was used as a postulate itself to
derive QM), and many more.  Dual to Postulate PFAITH an analogous postulate for effects would give
additional quantum features, such as teleportation. However, all possible consequences of these
postulates still need to be investigated, and it is not clear yet if one can derive QM from these
principles only.

\medskip In order to provide a route for seeking new candidates for operational postulates one can
short-circuit the axiomatic framework to select QM using a mathematical postulate dictated by what
is really special about the quantum theory, namely that not only transformations, but also {\em
  effects form a C${}^*$-algebra} (more precisely, this is true for all hybrid quantum-classical
theories, {\it i.e.} corresponding to QM plus super-selection rules). However, whereas the sum of effects
can be operationally defined, their composition has no operational meaning, since the notion itself
of ``effect'' abhors any kind of composition.  I will then show that with another natural postulate,
\medskip
\begin{itemize}
\item[AE:] atomicity of evolution,
\end{itemize}
\medskip
along with the mathematical postulate,
\medskip
\begin{itemize}
\item[CJ:] Choi-Jamiolkowski isomorphism \cite{choi75,Jamiolkowski72},
\end{itemize} 
\medskip
\par\noindent {\em it is possible to identify effects with ``atomic'' events}, {\it i.e.} elementary
events which cannot be refined as the union of events. Via the composition of atomic events we can
then define the composition of effects, thus selecting the quantum-classical hybrid among all
possible general probabilistic theories (including the PR-boxes, which indeed satisfy both NSF and
PFAITH).

The CJ isomorphism looks natural in an operational context, and it is hoped that it will be
converted soon into an operational postulate.

\bigskip
The present operational axiomatization will adhere to the following three general principles:
\medskip 
\begin{enumerate}
\item {\bf (Strong Copenhagen)} Everything is defined operationally, including all mathematical
  objects. Operationally indistinguishable entities are identified.
\item {\bf (Mathematical closure)} Mathematical completion is taken for convenience. 
\item {\bf (Operational closure)} Every operational option that is implicit in the formulation is
  incorporated in the axiomatic framework.
\end{enumerate}
\medskip An example of the application of the Strongly Copenhagen principle is the notion of {\em
  system}, which here I will identify with a collection of tests---the tests that can be performed
over the system.  A typical case of operational identification is that of events occurring with the
same probability and producing the same conditioning. Another case is the statement that the set of
states is separating for effects and viceversa.  Examples of mathematical closure are the norm
closure, the algebraic closure, and the linear span.  It is unquestionable that these are always
idealizations of operational limiting cases, or they are introduced just to simplify the
mathematical formulation ({\it e.g.} real numbers versus the ``operational'' rational numbers). Operational 
completeness, on the other hand, does not affect the corresponding mathematical representation, since
every incorporated option is already implicit in the formulation. This is the case, for example, for
convex closure, closure under coarse-graining, etc. which are already implicit in the probabilistic
formulation.

\section{C${}^*$-algebra representation of probabilistic theories}
\subsection{Tests and states}
A probabilistic operational framework is a collection of {\bf tests}\footnote{The present notion of
  test corresponds to that of {\bf experiment} of Ref.  \cite{darianoVax2006}. Quoted from the same
  reference: ``An experiment on an object system consists in making it interact with an apparatus,
  which will produce one of a set of possible events, each one occurring with some probability. The
  probabilistic setting is dictated by the need of experimenting with partial {\em a priori}
  knowledge about the system (and the apparatus).  In the logic of performing experiments to predict
  results of forthcoming experiments in similar preparations, the information gathered in an
  experiment will concern whatever kind of information is needed to make predictions, and this, by
  definition is the {\em state} of the object system at the beginning of the experiment. Such
  information is gained from the knowledge of which transformation occurred, which is the
  ``outcome'' signaled by the apparatus.''}  $\AA,\AB,\AC,\ldots$ each being a complete collection
$\AA=\{\tA_i\}$, $\AB=\{\tB_j\}$, $\AC=\{\tC_k\}$, \ldots of mutually exclusive {\bf events}
$\tA_i,\tB_j,\tC_k,\ldots$ occurring probabilistically\footnote{Also A.  R\'enyi \cite{Renyi97}
  calls our test ``experiment''.  More precisely, he defines an experiment $\AA$ as the pair
  $\AA=({\mathfrak X},\alg{A})$ made of the {\em basic space} ${\mathfrak X}$---the collection of
  outcomes---and of the $\sigma$-algebra of events $\alg{A}$. Here, to decrease the mathematical
  load of the framework, we conveniently identify the experiment with the basic space only, and
  consider a different $\sigma$-algebra ({\it e.g.} a coarse graining) as a new test made of new
  mutually exclusive events. Indeed, since we are considering only discrete basic spaces, we can put
  basic space and $\sigma$-algebra in one-to-one correspondence, by taking $\alg{A}=2^{{\mathfrak
      X}}$---the power set of ${\mathfrak X}$---and, viceversa, $\mathfrak X$ as the collection of the
  minimal intersections of elements of $\alg{A}$.}; events that are mutually exclusive are often
called {\bf outcomes}. The same event can occur in different tests, with occurrence probability
independent on the test.  A {\bf singleton test}---also called a {\bf channel}---$\AD=\{\tD\}$ is
{\bf deterministic}: it represents a non-test, {\it i.e.} a free evolution.  The {\bf union}
$\tA\cup\tB$ of two events corresponds to the event in which either $\tA$ or $\tB$ occurred, but it
is unknown which one. A {\bf refinement} of an event $\tA$ is a set of events $\{\tA_i\}$ occurring
in some test such that $\tA=\cup_i \tA_i$.  The experiment itself $\AA$ can be regarded as the
deterministic event corresponding to the complete union of its outcomes, and when regarded as an
event it will be denoted by the different notation $\tD_\AA$. The opposite event of $\tA$ in $\AA$
will be denoted as $\overline{\tA}:=\complement_\AA\tA$.\footnote{By adding the intersection of
  events, one builds up the full {\em Boolean algebra of events} (see {\it e.g.} Ref.
  \cite{Renyi97}).}  The union of events transforms a test $\AA$ into a new test $\AA'$ which is a
{\bf coarse-graining} of $\AA$, {\it e.g.}  $\AA=\{\tA_1,\tA_2,\tA_3\}$ and
$\AA'=\{\tA_1,\tA_2\cup\tA_3\}$. Vice-versa, we will call $\AA$ a {\bf refinement} of $\AA'$.

The {\bf state} $\omega$ describing the preparation of the system is the probability rule
$\omega(\tA)$ for any event $\tA\in\AA$ occurring in any possible test $\AA$.\footnote{By definition
  the state is the collection of the variables of a system whose knowledge is sufficient to make
  predictions.  In the present context, it allows one to predict the results of tests, whence it is
  the probability rule for all events in any conceivable test.}  For each test $\AA$ we have the
completeness $\sum_{\tA_j\in\AA}\omega(\tA_j)=1$. States themselves are considered as special tests:
the {\bf state-preparations}. 

\subsection{Cascading, conditioning, and transformations\label{s:NSF}}

The {\bf cascade} $\AB\circ\AA$ of two tests $\AA=\{\tA_i\}$ and $\AB=\{\tB_j\}$ is the new test
with events $\AB\circ\AA=\{\tB_j\circ\tA_i\}$, where $\tB\circ\tA$ denotes the {\bf composite event}
$\tA$ ``followed by'' $\tB$ satisfying the following \Postulate{NSF}{No-signaling from the future}{The
  marginal probability $\sum_{\tB_j\in\AB}\omega(\tB_j\circ\tA)$ of any event $\tA$ is independent
  on test $\AB$, and is equal to the probability with no test $\AB$, namely
\begin{equation}\label{e:NSF}
\sum_{\tB_j\in\AB}\omega(\tB_j\circ\tA)=:f(\AB,\tA)\equiv\omega(\tA),\quad\forall\AB,\,\tA,\,\omega.
\end{equation}}
NSF is part of the definition itself of test-cascade, however, we treat it as a
separate postulate, since it corresponds to the {\bf choice of the arrow of time}.\footnote{Postulate NSF is not just a Kolmogorov consistency
  condition for marginals of a joint probability. In fact, even though the marginal over test $\AB$
  in Eq. (\ref{e:NSF}) is obviously the probability of $\tA$, such probability in principle depends
  on the test $\AB$, since the joint probability generally depends on it. And, indeed, the
  marginal over entry $\tA$ does generally depend on the past test $\AA\ni\tA$. Such asymmetry of
  the joint probability under marginalization over future or past tests represents {\em the choice
    of the arrow of time}. Of course one could have assumed the opposite postulate of no-signaling
  from the past, considering conditioning from the future instead, thus reversing the arrow of time.
  Postulate NSF introduces conditioning from tests, and is part of the definition itself of temporal
  cascade-tests. The need of considering NSF as a Postulate has been noticed for the first time by
  Masanao Ozawa (private communication).} The interpretation of the test-cascade
$\AB\circ\AA$ is that ``test $\AA$ can influence test $\AB$ but not vice-versa.''\footnote{One could also
  defined more general cascades not in time, {\it e.g.} the circuit diagram\epsfig{file=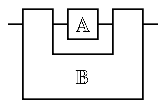,width=1.5cm}. 
  This would have given rise to a probabilistic version of the quantum comb theory of Ref.
  \cite{Chiribella:2008p3159}.} Postulate NSF allows one to define the conditioned probability
$p(\tB|\tA)=\omega(\tB\circ\tA)/\omega(\tA)$ of event $\tB$ occurring conditionally on the previous
occurrence of event $\tA$. It also guarantees that the probability of $\tB$ remains independent of
the test $\AB$ when conditioned.

Conditioning sets a new probability rule corresponding to the notion of {\bf conditional state}
$\omega_\tA$, which gives the probability that an event occurs knowing that event $\tA$ has occurred
with the system prepared in the state $\omega$, namely
$\omega_\tA\doteq\omega(\cdot\circ\tA)/\omega(\tA)$\footnote{Throughout the paper the central dot
  ``$\cdot$'' denotes the location of the pertinent variable.}. We can now regard the event $\tA$ as
transforming with probability $\omega(\tA)$ the state $\omega$ to the (unnormalized) state
$\tA\omega$\footnote{This is the same as the notion of {\em quantum operation} in QM, which gives
  the conditioning $\omega_\tA=\tA\omega/\tA\omega(\tI)$, or, in other words, the analogous of the
  quantum Schr\"{o}dinger picture evolution of states.}  given by
\begin{equation}
\tA\omega:=\omega(\cdot\circ\tA).
\end{equation}
Therefore, the notion of cascade and Postulate NSF entail the identification:

\bigskip \centerline{{\bf event}$\equiv${\bf transformation},}\bigskip

\noindent which in turn implies the equivalence:

\bigskip \centerline{{\bf evolution}$\equiv${\bf state-conditioning}.\footnote{Clearly this includes
    the deterministic singleton-test $\AD=\{\tD\}$---the analogues of quantum channels, including
    unitary evolutions.}}  \bigskip

Notice that operationally a transformation $\tA$ is completely specified by all the joint probabilities in
which it is involved, whence, it is univocally given by the probability rule
$\tA\omega=\omega(\cdot\circ\tA)$ for all states $\omega$. This is equivalent to specify both the
conditional state $\omega_\tA$ and the probability $\omega(\tA)$ for all possible states $\omega$,
due to the identity
\begin{equation}\label{completespec}
\tA\omega=\omega(\tA)\omega_\tA.
\end{equation}
In particular the {\bf identity transformation} $\tI$ is completely specified by the rule
$\tI\omega=\omega$ for all states $\omega$. 
\subsection{System}
In a pure Copenhagen spirit we will identify a {\bf system} $\System$ with a collection of {\bf
  tests} $\System=\{\AA,\AB,\AC,\ldots\}$, the collection being operationally closed under
coarse-graining, convex combination, conditioning, and cascading, and will includes all states as
special tests. Closure under cascading is equivalent to consider {\bf mono-systemic evolution}, {\it
  i.e.}  in which there are only tests for which the output system is the same of the input
one.\footnote{We could have considered more generally tests in which the output system is different
  from the input one, in which case the system is no longer closed under test-cascade, and, instead,
  there are cascades of tests from different systems. This would give more flexibility to the
  axiomatic approach, and could be useful for proving some theorems related to multipartite systems
  made of different systems. The fact that there are different systems would impose constraints to
  the cascades of tests, corresponding to allow only some particular words made of the ``alphabet''
  $\AA,\AB,\ldots$ of tests, and the system would then correspond to a ``language'' (see
  Ref.\cite{Chiribella:2009unp} for a similar framework).  Such generalization will be thoroughly
  analyzed in a forthcoming publication.\label{f:diffS} } The operator has always the option of
performing repeated tests, along with (randomly) alternating tests---say $\AA$ and $\AB$---in
different proportions---say $p$ and $1-p$ ($0<p< 1$)---thus achieving the test $\AC_p=p\AA+(1-p)\AB$
which is the {\bf convex combination} of tests $\AA$ and $\AB$, and is given by
$\AC_p=\{p\tA_1,p\tA_2,\ldots,(1-p)\tB_1,(1-p)\tB_2,\ldots\}$, where $p\tA$ is the same event as
$\tA$, but occurring with a probability rescaled by $p$. Since we will consider always the closure
under all operator's options (this is our {\em operational closure}), we will take the system also
to be closed under such convex combination.  In particular, the set of all states of the
system\footnote{At this stage such set not necessarily contains all {\em in-principle} possible
  states. The extension will be done later, after defining effects.} is
closed under convex combinations and under conditioning, and we will denote by $\Stset(\System)$
($\Stset$ for short) the convex set of all possible states of system $\System$. We will often use
the colloquialism ``for all possible states $\omega$'' meaning $\forall\omega\in\Stset(\System)$,
and we will do similarly for other operational objects.

In the following we will denote the set of all possible transformations/events by
$\Trnset(\System)$, $\Trnset$ for short. The convex
structure of $\System$ entails a convex structure for $\Trnset$, whereas the cascade of tests
entails the composition of transformations. The latter, along with the existence of the identity
transformation $\tI$, gives to $\Trnset$ the structure of {\em convex monoid}.

\subsection{Effects\label{s:effect}}
From the notion of conditional state two complementary types of equivalences for transformations
follow: the {\em conditional} and the {\em probabilistic} equivalence. The transformations $\tA_1$
and $\tA_2$ are {\bf conditioning-equivalent} when $\omega_{\tA_1}=\omega_{\tA_2}$
$\forall\omega\in\Stset$, namely when they produce the same conditional state for all prior states
$\omega$. On the other hand, the transformations $\tA_1$ and $\tA_2$ are {\bf probabilistically
  equivalent} when $\omega(\tA_1)=\omega(\tA_2)$ $\forall\omega\in\Stset$, namely when they occur
with the same probability.\footnote{In the previous papers
  \cite{dariano-losini2005,darianoVax2005,darianoVax2006,darianoVax2007} I called the conditional
  equivalence {\em dynamical equivalence}---since the two transformation will affect the same state
  change. However, one should more properly regard the ``dynamic'' change of the state $\omega$ due
  to the transformation $\tA$ as the unnormalized state $\tA\omega$, but the two transformations
  $\tA$ and $\tB$ will be fully equivalent when $\tA\omega=\tB\omega$ for all states $\omega$.
  Moreover, in the same previous papers I called the probabilistic equivalence {\em informational
    equivalence}---since the two transformations will give the same information about the state. The
  new nomenclature has a more immediate meaning.} Since operationally a transformation $\tA$ is
completely specified by the probability rule $\tA\omega$ for all states, it follows that two
transformations $\tA_1$ and $\tA_2$ are fully {\bf equivalent} ({\it i.e.} operationally indistinguishable)
when $\tA_1\omega=\tA_2\omega$ for all states $\omega$. We will identify two equivalent
transformations, and denote the equivalence simply as $\tA_1=\tA_2$. From identity
(\ref{completespec}) it follows that two transformations are equivalent if and only if they are both
conditioning and probabilistically equivalent.

A probabilistic equivalence class of transformations defines an {\bf effect}\footnote{This is the
  same notion of ``effect'' introduced by Ludwig \cite{Ludwig-axI}}. In the following we will denote
effects with lowercase letters $a,b,c,\ldots$ and denote by $[\tA]_\eff$ the effect containing
transformation $\tA$. We will also write $\tA\in a$ meaning that ``the transformation $\tA$ belongs
to the equivalence class $a$'', or ``$\tA$ has effect $a$'', and write ``$\tA\in[\tB]_\eff$ to say
that ``$\tA$ is probabilistically equivalent to $\tB$''. Since by definition
$\omega(\tA)=\omega([\tA]_\eff)$, hereafter we will legitimately write the variable of the state as
an effect, {\it e.g.} $\omega(a)$.  The {\bf deterministic effect} will be denoted by $e$, corresponding to
$\omega(e)=1$ for all states $\omega$. We will denote the set of effects for a system $\System$ as
$\Cntset(\System)$, or just $\Cntset$ for short. The set of effects inherits a convex structure from
that of transformations. 

By the same definition of state---as probability rule for transformations---states are separated by
effects (whence also by transformations\footnote{In fact, $\tA\omega\neq\tA\zeta$ for
  $\tA\in\Trnset$ means that there exists an effect $c$ such that $\tA\omega(c)\neq\tA\zeta(c)$,
  whence the effect $c\circ\tA$ will separates the same states.}), and, conversely, effects are
separated by states.  Transformations are separated by states in the sense that $\tA\neq\tB$ iff
$\tA\omega\neq\tB\omega$ for some state. As a consequence, it may happen that the introduction of a
new state via some new preparation (such as introducing additional systems) will separate two
previously indiscriminable transformations, in which case we will include the new state (and all
convex combination with it) in $\Stset(\System)$, and we will complete the system $\System$
accordingly.  We will end with $\Stset(\System)$ separating $\Trnset(\System)$ and
$\Cntset(\System)$, and $\Cntset(\System)$ separating $\Stset(\System)$.

The identity $\omega_\tA(\tB)\equiv \omega_\tA([\tB]_\eff)$ implies that
$\omega(\tB\circ\tA)=\omega([\tB]_\eff\circ\tA)$ for all states $\omega$, leading to the chaining rule
$[\tB]_\eff\circ\tA=[\tB\circ\tA]_\eff$, corresponding to the ``Heisenberg picture'' evolution in terms
of transformations acting on effects. Notice that transformations act on effects from the right,
inheriting the composition rule of transformations ($\tB\circ\tA$ means ''$\tA\text{ followed by
}\tB$'').  Notice also that $e\circ\tA\in[\tI\circ\tA]_\eff=a$. It follows that for $\tD$ deterministic
one has $\tD\in e$, whence $\tD\circ\tA\in[\tA]_\eff$.

Consistently, in the ``Schr\"{o}dinger picture'', we have
$\tB\omega(\cdot\circ\tA)=\omega(\cdot\circ\tB\circ\tA)$, corresponding to
$(\tB\circ\tA)\omega=\omega(\cdot\circ\tB\circ\tA)$. Also, we will use the unambiguous notation
$\tB\omega(a)=[\tB\omega](a)$, whence 
$\tB\omega(a)=\omega(a\circ\tB)$, and $\omega(a)=\tA\omega(e),\;\forall\tA\in a$.

\subsection{Linear structures for transformations and effects}\label{s:lineareffects}
Transformations $\tA_1$ and $\tA_2$, for which one has the bound $\omega(\tA_1)+\omega(\tA_2)\leq
1,\;\forall\omega\in\Stset$ can in principle occur in the same test, and we will call them {\bf
  test-compatible}. For test-compatible transformations one can define their addition $\tA_1+\tA_2$
via the probability rule
\begin{equation}\label{eq:sum}
(\tA_1+\tA_2)\omega=\tA_1\omega+\tA_2\omega, 
\end{equation}
were we remind that $\tA\omega:=\omega(\cdot\circ\tA)$. Therefore the sum of two test-compatible
transformations is just the union-event $\tA_1+\tA_2=\tA_1\cup\tA_2$, with the two transformations
regarded as belonging to the same test.\footnote{The probabilistic class of $\tA_1+\tA_2$ is given
  by $$\omega(\tA_1+\tA_2)=\omega(\tA_1)+\omega(\tA_2),\quad \forall\omega\in\Stset,$$ whereas the
 conditional class is given by $$\omega_{\tA_1+\tA_2}=\frac{\omega(\tA_1)}{\omega(\tA_1+\tA_2)} 
\omega_{\tA_1}+\frac{\omega(\tA_2)}{{\omega(\tA_1+\tA_2)}}\omega_{\tA_2},\quad
\forall\omega\in\Stset.$$} For any test $\AA$ we can define its {\bf total coarse-graining} as the
deterministic transformation $\tD_\AA=\sum_{\tA_i\in\AA}\tA_i$.
We can trivially extend the addition rule (\ref{eq:sum}) to any set of (generally non
test-compatible) transformations, and to subtraction of transformations as well. Notice that the composition
``$\circ$'' is distributive with respect to addition ``$+$''. 

We can define the multiplication $\lambda\tA$ of a transformation $\tA$ by a scalar
$0\leq\lambda\leq 1$ by the rule
\begin{equation}\label{eq:lambda}
\omega(\cdot\circ\lambda\tA) =\lambda\omega(\cdot\circ\tA),
\end{equation}
namely $\lambda\tA$ is the transformation conditioning-equivalent to $\tA$, but occurring with
rescaled probability $\omega(\lambda\tA)=\lambda\omega(\tA)$---as happens in the convex combination
of tests. It follows that for every couple of transformations $\tA$ and $\tB$ the transformations
$\lambda\tA$ and $(1-\lambda)\tB$ are test-compatible for $0\leq\lambda\leq 1$, consistently with
the convex closure of the system $\System$. By extending the definition (\ref{eq:lambda}) to any
positive $\lambda$, we then introduce the cone $\Trnset_+$ of transformations.  We will call an
event $\tA$ {\bf atomic} if it has no nontrivial refinement in any test, namely if it cannot be
written as $\tA=\sum_i\tA_i$ with $\tA_i\neq\lambda_i\tA$ for some $i$ and $0<\lambda_i<1$. Notice
that the identity transformation is not necessarily atomic.\footnote{For example, the identity
  transformation is refinable in classical abelian probabilistic theory, where states are of the
  form $\varrho=\sum_lp_l|l\>\< l|$, with $\{|l\>\}$ a complete orthonormal basis and $\{p_l\}$ a
  probability distribution. Here the identity transformation is given by $\tI=\sum_k|k\>\<
  k|\cdot|k\>\< k|$, $\{|k\>\}$, which is refinable into rank-one projection
  maps.\label{f:refineid}} The set of extremal rays of the cone $\Trnset_+$---denoted by
$\Erays(\Trnset_+)$---contains the atomic transformations.

The notions of i) test-compatibility, ii) sum, and iii) multiplication by a scalar, are naturally
inherited from transformations to effects via probabilistic equivalence, and then to states via
duality. Correspondingly, we introduce the cone of effects $\Cntset_+$, and, by duality, we extend
the cone of states $\Stset_+$ to the dual cone of $\Cntset_+$, completing the set of states $\Stset$
to the cone-base of $\Stset_+$ made of all positive linear functionals over $\Cntset_+$ normalized
at the deterministic effect, namely all in-principle legitimate states (in parallel we complete the
system $\System$ with the corresponding state-preparations). We call such a completion of the set of
states the {\bf no-restriction hypothesis for states}, corresponding to the {\bf state--effect
  duality}, namely the convex cones of states $\Stset_+$ and of effects $\Cntset_+$ are dual each
other.\footnote{In infinite dimensions one also takes the closure of cones.}  The state cone
$\Stset_+$ introduce a natural {\bf partial ordering} $\geq$ over states and over effect (via
duality), and one has $a\in\Cntset$ iff $0\leq a\leq e$. Thus the convex set $\Cntset$ {\bf is a
  truncation of the cone} $\Cntset_+$, whereas {\bf $\Stset$ is a base for the cone
  $\Stset_+$}\footnote{We remind the reader that a set $\set{B}\subset\set{C}$ of a cone $\set{C}$ in a vector
  space $\set{V}$ is called {\em base} of $\set{C}$ if $0\not\in\set{B}$ and for every point $u\in
  \set{C}, u\neq 0$, there is a unique representation $u=\lambda v$, with $v\in \set{B}$ and
  $\lambda >0$. Then, one has that $u\in\set{C}$ spans an extreme ray of $\set{C}$ iff $u=\lambda
  v$, where $\lambda>0$ and $v$ is an extreme point of $\set{B}$ (see Ref.\cite{Barvinok:2002}).}
defined by the normalization condition $\omega\in\Stset$ iff $\omega\in\Stset_+$ and $\omega(e)=1$.
In the following it will be useful also to express the probability rule $\omega(a)$ also in its dual form
$a(\omega)=\omega(a)$, with the effect acting on the state as a linear functional.

By extending to any real (complex) scalar $\lambda$ Eq. (\ref{eq:lambda}) we build the linear real
(complex) span $\Trnset_\Reals=\Span_\Reals(\Trnset)$ ($\Trnset_\Cmplx=\Span_\Cmplx(\Trnset)$).  The
{\em Cartesian decomposition} $\Trnset_\Cmplx=\Trnset_\Reals\oplus i\Trnset_\Reals$ holds, {\it i.e.} each
element $\tA\in\Trnset_\Cmplx$ can be uniquely written as $\tA=\tA_R+i\tA_I$, with
$\tA_R,\tA_I\in\Trnset_\Reals$.\footnote{Note that the elements $\tT\in\Trnset_\Reals$ can in turn
  be decomposed \`a la Jordan as $\tT=\tT_+-\tT_-$, with $\tT_\pm\in\Trnset_+$. However, such
  a decomposition is generally not unique. According to a theorem of B\'ellissard and Jochum
  \cite{Bellissard:1978p2020} the Jordan decomposition of the elements of the real span of a cone
  (with $\tT_\pm$ orthogonal in $\Trnset_\Reals$ Euclidean space) is unique if and only if the cone
  is self-dual.} Analogously, also for effects and states we define $\Cntset_\Field,\Stset_\Field$
for $\Field=\Reals,\Cmplx$.  The state--effect duality implies the linear space identifications
$\Stset_\Field\equiv\Cntset_\Field$.  Thanks to such identifications and to the identity of the
dimension of a convex cone with that of its complex and real spans, in the following without
ambiguity, we will simply write
$\dim(\System):=\dim[\Stset_+(\System)]\equiv\dim[\Cntset_+(\System)]$.  Moreover, if there is no
confusion, with some abuse of terminology we will simply refer by ``states,'' ``effects,'' and
``transformations'' to the respective generalized versions that are elements of the cones, or of their
real and complex linear spans.

Note that the cones of states and effects contain the origin, {\it i.e.} the null vector of the linear
space.  For the cone of states one has that $\omega=0$ iff $\omega(e)=0$ (since for any effect $a$
one has $0\leq\omega(a)\leq\omega(e)=0$, namely $\omega(a)=0$). On the other hand, the hyperplane
which truncates the cone of effects giving the physical convex set $\Cntset$ is conveniently
characterized using any {\bf internal state} $\vartheta$---{\it i.e.} a state that can be written as the
convex combination of any state with some other state---by using the following lemma
\begin{lemma} For any $a\in\Cntset_+$ one has $a=0$ iff $\vartheta(a)=0$ and $a=e$  iff
  $\vartheta(a)=1$, with $\vartheta$ any internal state.
\end{lemma}
\Proof For any state $\omega$ one can write $\vartheta=p\omega+(1-p)\omega'$ with $0\leq p\leq 1$
and $\omega'\in\Stset$. Then one has $\vartheta(a)=0$ iff $\omega(a)=0$ $\forall\omega\in\Stset$, that is 
iff $a=0$. Moreover, one has $\vartheta(a)=1$ iff $\omega(a)=1$ $\forall\omega\in\Stset$, {\it i.e.}
$a=e$. \qed

\subsection{Observables and informational completeness}
An {\bf observable} $\AL$ is a complete set of effects $\AL=\{l_i\}$ summing to the deterministic
effect as $\sum_{l_i\in\AL}l_i=e$, namely $l_i$ are the effects of the events of a test.  An
observable $\AL=\{l_i\}$ is named {\bf informationally complete} for $\System$ when each effect can
be written as a real linear combination of $l_i$, namely
$\Span_\Reals(\AL)=\Cntset_\Reals(\System)$. When the effects of $\AL$ are linearly
independent the informationally complete observable is named {\em minimal}.  Clearly, since
$\Cntset$ is separating for states, {\bf any informationally complete observable separates states},
that is using an informationally complete observable we can reconstruct also any state
$\omega\in\Stset(\System)$ from the set of probabilities $\omega(l_i)$. The existence of a minimal
informationally complete observable constructed from the set of available tests is guaranteed by the
following Theorem:
\begin{theorem}[Existence of minimal informationally complete observable]\label{th:infocomplete}
  It is always possible to construct a minimal informationally complete observable for $\System$ out
  of a set of tests of $\System$.
\end{theorem}
For the proof see Ref. \cite{darianoVax2007}.

\bigskip
\par In the following we will take a fixed minimal informationally complete observable $\AL=\{l_i\}$
as a {\bf reference test}, with respect to which all basis-dependent representations will be
defined.

\bigskip Symmetrically to the notion of informationally complete observable we have the notion of {\bf
  separating set of states} $\AS=\{\omega_i\}$, in terms of which one can write any state as a real
linear combination of the states $\{\omega_i\}$, namely $\Stset_\Reals(\System)=\Span_\Reals(\AS)$.
Regarded as a test $\AS=\{\tS_i\}\in\System$ the set of states $\{\omega_i\}$ correspond to the
state-reduction $\tS_i\omega=\omega(\tS_i)\omega_i$, $\forall\omega\in\Stset$. When the
corresponding effects $[\tS_i]_\eff$ form an informationally complete observable the test $\AS$
would be an example of the {\em Quantum Bureau International des Poids et Mesures} of Fuchs
\cite{Fuchs:2002p89}.

\subsection{Banach structures\label{s:Banach}}
On states $\omega\in\Stset$ introduce the {\bf natural norm}
$\n{\omega}=\sup_{a\in\Cntset}\omega(a)$, which extends to the whole linear space $\Stset_\Reals$ as
$\n{\omega}=\sup_{a\in\Cntset}|\omega(a)|$. Then, we can introduce the dual norm on effects
$\n{a}:=\sup_{\omega\in\Stset_\Reals,\n{\omega}\leq 1}|\omega(a)|$, and then on transformations
$\n{\tA}:=\sup_{b\in\Cntset_\Reals,\n{b}\leq 1}\n{b\circ\tA}$. Closures in norm (for mathematical
convenience) make $\Cntset_\Reals$ and $\Stset_\Reals$ a dual Banach pair, and $\Trnset_\Reals$ a
real Banach algebra.\footnote{An algebra of maps over a Banach space inherits the norm induced by
  that of the Banach space on which it acts. It is then easy to prove that the closure of the
  algebra under such norm is a Banach algebra.} Therefore, all operational quantities can be
mathematically represented as elements of such Banach spaces.

\subsection{Metric}
One can define a {\bf natural distance} between states $\omega,\zeta\in\Stset$ as follows
\begin{equation}
d(\omega,\zeta):=\sup_{l\in\Cntset}l(\omega)-l(\zeta).\label{state-dist}
\end{equation}
\begin{lemma}\label{l:metric} The function (\ref{state-dist}) is a metric on $\Stset$, and is
  bounded as $0\leq d(\omega,\zeta)\leq 1$. 
\end{lemma}
\Proof For every effect $l$, $e-l$ is also a effect, whence
\begin{equation}\label{ndist}
\begin{split}
d(\omega,\zeta)=&\sup_{l\in\Cntset}(l(\omega)-l(\zeta))=
\sup_{l'\in\Cntset}((e-l')(\omega)-(e-l')(\zeta))\\=&
\sup_{l'\in\Cntset}(l'(\zeta)-l'(\omega))=d(\zeta,\omega),
\end{split}
\end{equation}
that is $d$ is symmetric. On the other hand, $d(\omega,\zeta)=0$ implies that $\zeta=\omega$, since
the two states must give the same probabilities for all transformations. Finally, one has
\begin{equation}
\begin{split}
d(\omega,\zeta)&=\sup_{l\in\Cntset}(l(\omega)-l(\theta)+l(\theta)-l(\zeta))\\&\le
\sup_{l\in\Cntset}(l(\omega)-l(\theta))+
\sup_{l\in\Cntset}(l(\theta)-l(\zeta))=d(\omega,\theta)+d(\theta,\zeta),
\end{split}
\end{equation}
that is it satisfies the triangular inequality, whence $d$ is a metric. By construction, the distance
is bounded as $d(\omega,\zeta)\leq 1$, since the maximum value of $d(\omega,\zeta)$ is achieved when
$l(\omega)=1$ and $l(\zeta)=0$.\qed

The natural distance (\ref{ndist}) is extended to a metric over $\Stset_\Reals$ as 
$d(\omega,\zeta)=\n{\omega-\zeta}$ with $\n{\cdot}$ the norm over $\Stset_\Reals$. Analogously we
define the distance between effects as $d(a,b):=\sup_{\omega\in\Stset}|\omega(a-b)|$.\footnote{It is
easy to check that such distance satisfies the trangular inequality}

\medskip
A relevant property of the metric in Eq. (\ref{state-dist}) is its {\bf monotonicity}, that
the distance between two states can never increase under deterministic evolution, as established
by the following lemma.

\begin{lemma}[Monotonicity of the state distance] For every deterministic physical transformation
  $\tD\in\Trnset$, one has
\begin{equation}
d(\tD\omega,\tD\zeta)\leq d(\omega,\zeta).
\end{equation}
\end{lemma}
\Proof First we notice that since $\tD\in\Trnset$ is a physical transformation, for every effect
$a\in\Cntset$ one has also $a\circ\tD\in\Cntset$, whence $\Cntset\circ\tD\subseteq\Cntset$.
Therefore, we have
\begin{equation}
\begin{split}
d(\tD\omega,\tD\zeta):=&\sup_{a\in\Cntset}\omega(a\circ\tD)-\zeta(a\circ\tD)\\=&
\sup_{a\in\Cntset\circ\tD}\omega(a)-\zeta(a)\leq\sup_{a\in\Cntset}\omega(a)-\zeta(a)=
d(\omega,\zeta).
\end{split}
\end{equation}
Notice that we take the transformation deterministic only to assure that $\tD\omega$ is itself a
state for any $\omega$.  
\qed

\subsection{Isometric transformations} A deterministic transformation $\tU$ is called {\em
  isometric} if it preserves the distance between states, namely
\begin{equation}
d(\tU\omega,\tU\zeta)\equiv d(\omega,\zeta),\qquad \forall \omega,\zeta\in\Stset.
\end{equation}
\begin{lemma} In finite dimensions, all the following properties of a transformation are equivalent:
  (a) it is isometric for $\Stset$; (b) it is isometric for $\Cntset$; (c) it is automorphism of
  $\Stset$; (d) it is automorphism of $\Cntset$.
\end{lemma}
\Proof By definition a transformation of the convex set (of states or effects) is a linear map of
the convex set in itself.  A linear isometric map of a set in itself is isometric on the linear span
of the set.\footnote{Interestingly, the Mazur-Ulam theorem states that any surjective isometry (not
  necessarily linear) between real normed spaces is affine. Therefore, even if nonlinear, it would
  map convex subsets to convex subsets.} (Recall that the natural distance between states has been
extended to a metric over the whole $\Stset_\Reals$.) In finite dimensions an isometry on a normed
linear space is diagonalizable \cite{Koehler:1970p3754}. Its eigenvalues must have unit modulus,
otherwise it would not be isometric. It follows that it is an orthogonal transformation, and since
it maps the set into itself, it must be a linear automorphism of the set. Therefore, an isometric
transformation of a convex set is an automorphism of the convex set\footnote{For a convex set, an
  automorphism must send the set to itself keeping the convex structure, whence it must be a
  one-to-one map that is linear on the span of the convex set.}

Now, automorphisms of $\Stset$ are isometric for $\Cntset$, since
\begin{equation}
\begin{split}
d(a\circ\tU,b\circ\tU)=&\sup_{\omega\in\Stset}|\omega((a-b)\circ\tU)|=
\sup_{\omega\in\Stset}|(\tU\omega)(a-b)|\\=&\sup_{\omega\in\tU\Stset}|\omega(a-b)|
=\sup_{\omega\in\Stset}|\omega(a-b)|=d(a,b),
\end{split}
\end{equation}
and, similarly, automorphisms of $\Cntset$ are isometric for $\Stset$, since
\begin{equation}
\sup_{a\in\Cntset}[\omega(a\circ\tU)-\zeta(a\circ\tU)]=\sup_{a\in\Cntset\circ\tU}[\omega(a)-\zeta(a)]=d(\omega,\zeta).
\end{equation}
Therefore, automorphisms of $\Stset$ are isometric for $\Cntset$, whence, for the first part of the
proof, they are automorphisms of $\Cntset$, whence they are isometric for $\Stset$.
\qed 

\medskip
The {\em physical automorphisms} play the role of unitary transformations in QM.
\begin{corollary}[Wigner theorem] The only transformations of states that are inverted by another
  transformation, must send pure states to pure states, and are isometric.
\end{corollary}
\subsection{The C${}^*$ algebra of transformations\label{s:Cstar}}
\begin{figure}[h]
\epsfig{file=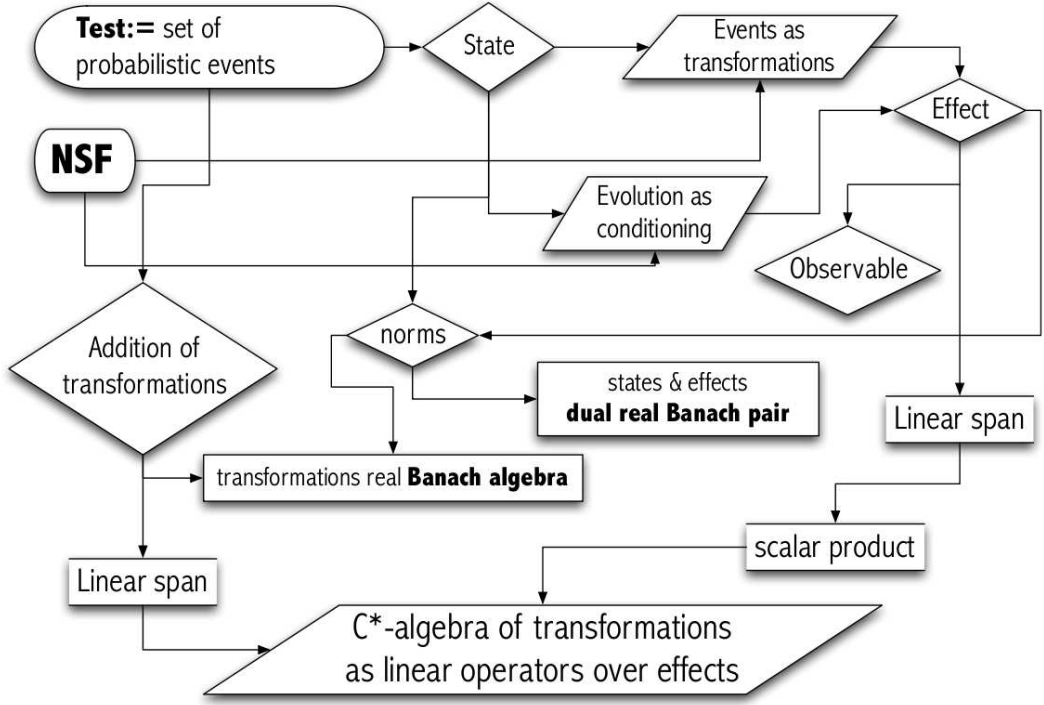,width=14cm}
\caption{Logical flow chart leading to the representation of any probabilistic theory in terms of a
  C${}^*$-algebra of linear transformations over the linear space of complex effects (see also
  Footnote \ref{f:faithscal} and Subsect. \ref{s:scalPhi} for an operational basis for the scalar
  product).}\label{fig:flowCstar}
\end{figure}

We can represent the transformations as elements of $\Trnset_\Cmplx$ regarded as a complex
C${}^*$-algebra. This is obvious, since $\Trnset_\Cmplx$ are by definition linear transformations of
effects, making an associative sub-algebra $\Trnset_\Cmplx\subseteq\Bnd{\Cntset_\Cmplx}$ of the
matrix algebra over $\Cntset_\Cmplx$. {\bf Adjoint} and {\bf norm} can be easily defined in terms of
any chosen {\bf scalar product $(\cdot,\cdot)$ over $\Cntset_\Cmplx$}, with the adjoint defined as
$(a\circ \tA^\dag,b)=(a,b\circ\tA)$, and the norm as
$\n{\tA}=\sup_{a\in\Cntset_\Cmplx}\n{a\circ\tA}/\n{a}$, with $\n{a}=\sqrt{(a,a)}$. (Notice that
these norms are different from the ``natural norms'' defined in Subsect. \ref{s:Banach}.) We can
then extend the complex linear space $\Trnset_\Cmplx$ by adding the adjoint transformations and
taking the norm-closure. We will denote such extension with the same symbol $\Trnset_\Cmplx$, which
is now a C${}^*$-algebra. Indeed, upon reconstructing $\Cntset_\Cmplx$ and $\Trnset_\Cmplx$ from the
original real spaces via the Cartesian decomposition $\Cntset_\Cmplx=\Cntset_\Reals\oplus
i\Cntset_\Reals$ and $\Trnset_\Cmplx=\Trnset_\Reals\oplus i\Trnset_\Reals$, and introducing the
scalar product on $\Cntset_\Cmplx$ as the sesquilinear extension of a real symmetric scalar product
$(\cdot,\cdot)_\Reals$ over $\Cntset_\Reals$, the adjoint of a real element $\tA\in\Trnset_\Reals$
is just the transposed matrix $\transp{\tA}$ with respect to a real basis orthonormal for
$(\cdot,\cdot)_\Reals$, and $\tA^\dag:=\transp{\tA_R}-i\transp{\tA_I}$ for a general
$\tA=\tA_R+i\tA_I\in\Trnset_\Cmplx$.  A natural choice of matrix representation for $\Trnset_\Reals$
is given by its action over a minimal informational complete observable $\AL=\{l_i\}$ (the scalar
product $(\cdot,\cdot)_\Reals:= (\cdot,\cdot)_\AL$ will correspond to declare $\AL$ as orthonormal).
Upon expanding $[l_i\circ\tA]_\eff$ again over $\AL=\{l_i\}$ one has the matrix representation
$l_i\circ\tA=\sum_{j}\tA_{ji}l_j$. Using the fact that $\AL$ is state-separating, we can write the
probability rule as the pairing $\omega(a)=(\omega,a)_\Reals$ between $\Cntset_\Reals$ and
$\Stset_\Reals$ (and analogously for their complex spans).\footnote{The present derivation of the
  C${}^*$-algebra representation of transformations is more direct than that in Ref.
  \cite{darianoVax2007}, and is just equivalent to the probabilistic framework inherent in the
  notion of ``test'' (see also the summary of the whole logical deduction in the flow chart in Fig.
  \ref{fig:flowCstar}). The specific C${}^*$-algebra in Ref.  \cite{darianoVax2007} possessed
  operational notions of adjoint and of scalar product over effects, both constructed using a
  symmetric faithful bipartite state, needing in this way the two additional postulates: a) the
  existence of dynamically independent systems, b) the existence of faithful symmetric bipartite
  states. Such construction is briefly reviewed in Subsect. \ref{s:scalPhi}.\label{f:faithscal}} In
this way we see that for every probabilistic theory one can always represent transformations/events
as elements of the C${}^*$-algebra $\Trnset_\Cmplx$ of matrices acting on the linear space of
complex effects $\Cntset_\Cmplx$. In Fig. \ref{fig:flowCstar} the logical derivation of the
C${}^*$-algebra representation of the theory is summarized.

Conversely, given (1) a C${}^*$-algebra $\Trnset_\Cmplx$, (2) the cone of transformations
$\Trnset_+$, and (3) the vector $e\in\Cntset_\Cmplx$ representing the deterministic effect, we can
rebuild the full probabilistic theory by constructing the cone of effects as the orbit  $\Cntset_+=e\circ\Trnset_+$, 
and taking the cone of states $\Stset_+$ as the dual cone of $\Cntset_+$.\footnote{The
  ``orbit'' $e\circ\Trnset_+$ is defined as the set:
  $e\circ\Trnset_+:=\{e\circ\tA|\tA\in\Trnset_+\}$.}
\section{Independent systems}
\subsection{Dynamical independence and marginal states\label{s:dynind}} 
A purely dynamical notion of {\em system independence} coincides with the possibility of performing
local tests. To be precise, we will call systems $\System_1$ and $\System_2$ {\bf independent} if it is
possible to perform their tests as {\bf local tests}, {\it i.e.} in such a way that for every joint state
of $\System_1$ and $\System_2$ the transformations on $\System_1$ commute with transformations on
$\System_2$, namely\footnote{The present definition of independent systems is purely dynamical, in
  the sense that it does not involve statistical requirements, {\it e.g.} the existence of factorized
  states. This, however, is implied by the mentioned no-restriction hypothesis for states.}
\begin{equation}
\tA^{(1)}\circ\tB^{(2)}=\tB^{(2)}\circ\tA^{(1)},\;\forall \tA^{(1)}\in\AA^{(1)},\,\forall
\tB^{(2)}\in\AB^{(2)}.
\end{equation}
The local tests comprise the Cartesian product $\System_1\times\System_2$, which is closed under
cascade. We will close this set also under convex combination, coarse-graining and conditioning,
making it a ``system'', and denote such a system with the same symbol $\System_1\times\System_2$, and
call {\bf local} all tests in $\System_1\times\System_2$. We now {\bf compose} the two systems
$\System_1$ and $\System_2$ into the {\bf bipartite} system $\System_1\odot\System_2$ by adding the
local tests into the new system $\System_1\odot\System_2$ as
$\System_1\odot\System_2\supseteq\System_1\times\System_2$ and closing under cascading,
coarse-graining and convex combination. We call the tests in
$\System_1\odot\System_2\setminus\System_1\times\System_2$ {\bf non-local}, and we will extend the
local/non-local nomenclature to the pertaining transformations.
In the following for identical systems we will also use the notation $\System^{\odot
  N}=\System\odot\System\odot\ldots\odot\System$ ($N$ times), and ${\mathfrak Z}^{\odot
  N}:={\mathfrak Z}(\System^{\odot N})$ to denote $N$-partite sets/spaces, with $\mathfrak
Z=\Stset,\Stset_+,\Stset_\Reals,\Stset_\Cmplx,\Cntset,\Cntset_+,\ldots$
 
Since the local transformations commute, we will just put them in a string, as
$(\tA,\tB,\tC,\ldots):= \tA^{(1)}\circ\tA^{(2)}\circ\tA^{(3)}\circ\ldots$ (convex combinations and
coarse graining will be sums of strings). Clearly, since the probability
$\omega(\tA,\tB,\tC,\ldots)$ is independent of the time ordering of transformations, it is just a
function only of the effects
$\omega(\tA,\tB,\tC,\ldots)=\omega([\tA]_\eff,[\tB]_\eff,[\tC]_\eff,\ldots)$, namely the joint
effect corresponding to local transformations is made of (sums of) local effects
$[(\tA,\tB,\tC,\ldots)]_\eff\equiv([\tA]_\eff,[\tB]_\eff,[\tC]_\eff,\ldots)$.

The embedding of local tests $\System_1\times\System_2$ into the bipartite system
$\System_1\odot\System_2$ implies that $\Trnset_\Field(\System_1\odot\System_2)\supseteq
\Trnset_\Field(\System_1)\otimes\Trnset_\Field(\System_2)$ and
$\Cntset_\Field(\System_1\odot\System_2)\supseteq
\Cntset_\Field(\System_1)\otimes\Cntset_\Field(\System_2)$, both for real and complex spans
$\Field=\Reals,\Cmplx$. On the other hand, since local tests include local state-preparation (or,
otherwise, because of the no-restriction hypothesis for states) the set of bipartite states
$\Stset(\System_1\odot\System_2)$ always includes the {\bf factorized states}, {\it i.e.} those
corresponding to factorized probability rules {\it e.g.} $\Omega(a,b)=\omega_1(a)\omega_2(b)$ for local
effects $a$ and $b$.  In parallel with local transformations and effects, we will denote factorized
states as strings $\Omega=(\omega_1,\omega_2,\ldots)$, {\it e.g.}
$(\omega_1,\omega_2)(a,b)=\omega_1(a)\omega_2(b)$.  Then, closure under convex combination implies
that $\Stset_\Field(\System_1\odot\System_2)\supseteq
\Stset_\Field(\System_1)\otimes\Stset_\Field(\System_2)$, for $\Field=\Reals,\Cmplx$.

For $N$ systems in the joint state $\Omega$, we define the {\bf marginal state} $\Omega|_n$ of the
$n$-th system the probability rule for any local transformation $\tA$ at the $n$-th system, with all
other systems untouched, namely
\begin{equation}\label{e:defloc}
\Omega|_n(\tA)\doteq\Omega(\tI,\ldots,\tI,\underbrace{\tA}_{n\text{th}},\tI,\ldots).
\end{equation}
Clearly, since the probability for local transformations depends only on their respective effects,
the marginal state is equivalently defined as
\begin{equation}
\Omega|_n(a)\doteq\Omega(e,\ldots,e,\underbrace{a}_{n\text{th}},e,\ldots)\quad\text{for } a\in\Cntset.
\end{equation}
It readily follows that the marginal state $\Omega|_n$ is independent of any deterministic
transformation---{\it i.e.} any test---that is performed on systems different from the $n$th: this is
exactly the general statement of the {\bf no-signaling} or {\bf acausality of local tests}.
Therefore, the present notion of dynamical independence directly implies no-signaling. The
definition in Eq.  (\ref{e:defloc}) can be trivially extended to unnormalized
states.\footnote{Notice that any generally unnormalized state is zero iff the joint state is zero,
  since $\Omega(e,e,\ldots,e)=\Omega_n(e)=0$.} $\!\!{}^,$\footnote{The present notion of dynamical
  independence is indeed so minimal that it can be satisfied not only by the quantum tensor product,
  but also by the quantum direct sum \cite{meta-quantum_trieste06}. (Notice, however, that an
  analogous of the Tsirelson's theorem \cite{Scholz:2008p3363} for transformations in finite
  dimensions would imply a representation of dynamical independence over the tensor-product of
  effects.)  In order to extract only the tensor product an additional assumption is needed. As
  shown in Refs.  \cite{meta-quantum_trieste06,darianoVax2007} two possibilities are either
  postulating the existence of bipartite states that are dynamically and preparationally faithful,
  or postulating the local observability principle. Here we will consider the former as a postulate,
  and derive the latter as a theorem. }

In the following we will use the following identities
\begin{equation}\label{id:marg}
\Psi|_2(a)=\Psi(e,a)=\Psi(e,e\circ\tA)=(\tI,\tA)\Psi(e,e),\;\forall\tA\in a.
\end{equation}

\begin{figure}[hbt]
\epsfig{file=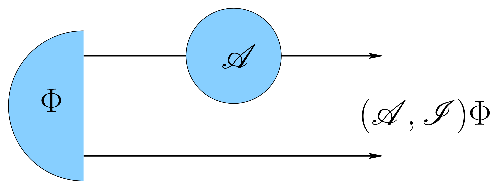,width=5cm}\hskip 2cm\epsfig{file=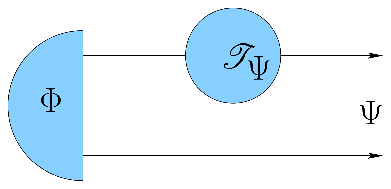,width=5cm} 
  \caption{Illustration of the notions of dynamically (left figure) and preparationally (right
    figure) faithful state for a bipartite system. A bipartite state $\Phi$ is dynamically faithful
    with respect to system $\System_1$ when the output state $(\tA,\tI)\Phi$ is in one-to-one
    correspondence with the local transformation $\tA$ on system $\System_1$, whereas it is
    preparationally faithful with respect to $\System_1$ if every bipartite state $\Psi$ can
    be achieved as $\Psi=(\tT_\Psi,\tI)\Phi$ via a local transformation $\tT_\Psi$ on $\System_1$.}
\end{figure}

\subsection{Faithful states} 
A bipartite state $\Phi\in\Stset(\System_1\odot\System_2)$ is {\bf dynamically faithful} with
respect to $\System_1$ when the output state $(\tA,\tI)\Phi$ is in one-to-one correspondence with
the local transformation $\tA$ on system $\System_1$, that is the cone-homomorphism\footnote{A
  cone-homomorphism between cones $\set{C}_1$ and $\set{C}_2$ is a linear map between
  $\Span_\Reals(\set{C}_1)$ and $\Span_\Reals(\set{C}_2)$ which sends elements of $\set{C}_1$ to
  elements of $\set{C}_2$, but not necessarily vice-versa.}  $\tA\leftrightarrow(\tA,\tI)\Phi$ from
$\Trnset_+(\System_1)$ to $\Stset_+(\System_1\odot\System_2)$ is a monomorphism.\footnote{This means
  that $(\tA_1,\tI)\Phi=(\tA_2,\tI)\Phi$ iff $\tA_1=\tA_2$, or, in other words,
  $\forall\tA\in\Trnset_\Reals$: $(\tA,\tI)\Phi=0\Longleftrightarrow\tA=0$.}  Equivalently the map
$\tA\mapsto(\tA,\tI)\Phi$ extends to an injective linear map between the linear spaces
$\Trnset_\Reals(\System_1)$ to $\Stset_\Reals(\System_1\odot\System_2)$ preserving the partial
ordering relative to the spanning cones, and this is true also in the inverse direction on the range
of the map. Notice that no physical transformation $\tA\neq 0$ ``annihilates'' $\Phi$, {\it i.e.} giving
$(\tA,\tI)\Phi=0$. 

\medskip A bipartite state $\Phi\in\Stset(\System_1\odot\System_2)$ is called {\bf preparationally
  faithful} with respect to $\System_1$ if every bipartite state $\Psi$ can be achieved as
$\Psi=(\tT_\Psi,\tI)\Phi$ by a local transformation $\tT_\Psi\in\Trnset_+(\System_1)$. This means
that the cone-homomorphism $\tA\mapsto(\tA,\tI)\Phi$ from $\Trnset_+(\System_1)$ to
$\Stset_+(\System_1\odot\System_2)$ is an epimorphism. Equivalently, the map
$\tA\mapsto(\tA,\tI)\Phi$ extends to a surjective linear map between the linear spaces
$\Trnset_\Reals(\System_1)$ to $\Stset_\Reals(\System_1\odot\System_2)$ preserving the partial
ordering relative to the spanning cones. 

\medskip In simple words, a dynamically faithful state keeps the imprinting of a local
transformation on the output, {\it i.e.} from the output we can recover the transformation. On the other
hand, a preparationally faithful state allows to prepare any desired joint state (probabilistically)
by means of local transformations. Dynamical and preparational faithfulness correspond to the
properties of being {\em separating} and {\em cyclic} for the C${}^*$-algebra of transformations.

\begin{theorem}\label{t:faith} The following assertions hold:
\begin{enumerate}
\item Any state $\Phi\in\Stset(\System_1\odot\System_2)$ that is preparationally faithful with
  respect to $\System_1$ is dynamically faithful with respect to $\System_2$.
\item\label{i:atomPhi} For identical systems in finite dimensions any state $\Phi$ that is
  preparationally faithful with respect to a system is also dynamically faithful with respect to the
  same system, and one has the cone-isomorphism\footnote{We say that two cones $\set{C}_1$ and
    $\set{C}_2$ are isomorphic (denoted as $\set{C}_1\simeq\set{C}_2$), if there exists a one-to-one
    linear mapping between $\Span_\Reals(\set{C}_1)$ and $\Span_\Reals(\set{C}_2)$ that is
    cone-preserving in both directions. We will call such a map a cone-isomorphism between the two
    cones.  Such a map will send extremal rays of $\set{C}_1$ to extremal rays of $\set{C}_2$, and
    positive linear combinations to positive linear combinations, and the same is true for the
    inverse map.\label{f:isocones}} $\Trnset_+(\System)\simeq\Stset_+(\System^{\odot 2})$. Moreover,
  a local transformation on $\Phi$ produces an output pure (unnormalized) bipartite state iff the
  transformation is atomic.
\item If there exists a state of $\System_1\odot\System_2$ that is preparationally faithful with
  respect to $\System_1$, then   $\dim(\System_1)\geq\dim(\System_2)$. 
\item\label{i:coneiso12} If there exists a state of $\System_1\odot\System_2$ that is
  preparationally faithful with respect to both systems, then one has the cone-isomorphisms
  $\Cntset_+(\System_1)\simeq\Stset_+(\System_2)$ and
  $\Cntset_+(\System_2)\simeq\Stset_+(\System_1)$.
\item If for two identical systems there exists a state that is preparationally faithful with
  respect to both systems, then one has the cone-isomorphism $\Stset_+\simeq\Cntset_+$ (weak
  self-duality).
\item If the state $\Phi\in\Stset(\System_1\odot\System_2)$ is preparationally faithful with respect
  to $\System_1$, for any invertible transformation $\tA\in\Trnset_+(\System_1)$ also the
  (unnormalized) state $(\tA,\tI)\Phi$ is preparationally faithful with respect to the same system.
  In particular, it will be a faithful state for any physical automorphism of
  $\Stset(\System_1)$.\footnote{One may be tempted to consider all automorphisms of
    $\Stset(\System_1)$, instead of just the physical ones. However, there is no guarantee that any
    automorphism will be also an automorphism of bipartite states when applied locally.  This is the
    case of QM, where the transposition is an automorphism of $\Stset(\System_1)$, nevertheless is
    not a local automorphism of $\Stset(\System_1\odot\System_2)$.\label{foot:auto}}
\item For identical systems in finite dimensions, for $\Phi$ preparationally faithful with respect
  to both systems, the state $\chi:=\Phi(e,\cdot)$ is cyclic in $\Stset_+(\System)$ under
  $\Trnset_+(\System)$, and the observables $\AL=\{l_i\}$ of $\System_2$ are in one-to-one
  correspondence with the ensemble decompositions $\{\rho_i\}_{i=1}^{|\AL|}$ of $\chi$, with
  $\rho_i:=\Phi(l_i,\cdot)$, and $\chi$ is an internal state.
\end{enumerate}
\end{theorem}
\Proof 
\begin{enumerate}
\item Introduce the map $\omega\mapsto\tT_{\omega}$ where for every $\omega\in\Stset(\System_2)$ one
  chooses a local transformation $\tT_{\omega}$ on $\System_1$ such that
  $(\tT_{\omega},\tI)\Phi|_2=\omega$. This is possible because $\Phi$ is preparationally faithful
  with respect to $\System_1$. One has
  $\tA\omega=(\tT_{\omega},\tA)\Phi|_2=(\tT_{\omega},\tI)(\tI,\tA)\Phi|_2$
  $\forall\omega\in\Stset(\System_2)$. Therefore, from $(\tI,\tA)\Phi$ one can recover the action of
  $\tA$ on any state $\omega$ by first applying $(\tT_{\omega},\tI)$ and then take the marginal, {\it i.e.}
  one recovers $\tA$ from $(\tI,\tA)\Phi$, which is another way of saying that
  $\tA\mapsto(\tI,\tA)\Phi$ is injective, namely $\Phi$ is dynamically faithful with respect to
  $\System_2$.

\item Denote by $\Phi\in\Stset^{\odot 2}$ a state that is preparationally faithful with respect to
  $\System_1$.  Since the linear map $\tA\mapsto(\tA,\tI)\Phi$ from $\Trnset_\Reals$ to
  $\Stset_\Reals^{\odot 2}$ is surjective, one has $\dim(\Trnset_\Reals)\geq
  \dim(\Stset_\Reals^{\odot 2})$.  However, one has also
  $\dim(\Trnset_\Reals)\leq\dim(\Stset_\Reals^{\odot 2})$ since $\Trnset_\Reals\subseteq
  \Bnd{\Stset_\Reals}\simeq\Stset_\Reals^{\otimes 2}\subseteq\Stset_\Reals^{\odot 2}$, whence
  $\dim(\Trnset_\Reals)=\dim(\Stset_\Reals^{\odot 2})$, and, having null kernel, the map is also
  injective, whence $\Phi$ is dynamically faithful with respect to $\System_1$. Since now the state
  $\Phi$ is both preparationally and dynamically faithful with respect to the same system
  $\System_1$, it follows that the map $\tA\mapsto(\tA,\tI)\Phi$ establishes the cone-isomorphism
  $\Trnset_+\simeq\Stset_+^{\odot 2}$. Since the faithful state establishes the cone-isomorphism
  $\Trnset_+\simeq\Stset_+^{\odot 2}$, it maps extremal rays of $\Trnset_+$ to extremal rays of
  $\Stset_+^{\odot 2}$ and vice-versa, that is $\tA\in\Erays(\Trnset_+)$ iff
  $(\tA,\tI)\Phi\in\Erays(\Stset_+^{\odot 2})$.

\item For $\Phi$ preparationally faithful with respect to $\System_1$, consider the cone homomorphism
$a\mapsto\omega_a:=\Phi(a,\cdot)$ which associates an (un-normalized) state
$\omega_a\in\Stset_+(\System_2)$ to each effect $a\in\Cntset_+(\System_1)$. The extension to a
linear map $a\mapsto\omega_a$ between the linear spaces $\Stset_\Reals(\System_2)$ and
$\Cntset_\Reals(\System_1)$ preserves the cone structure, and is surjective, since $\Phi$ is
preparationally faithful with respect to $\System_1$ (whence every bipartite state, and in particular
every marginal state, can be obtained from a local effect). The bound
$\dim(\System_1)\geq\dim(\System_2)$ then follows from surjectivity.
\item Similarly to the proof of item (1), consider the map $\lambda\mapsto\tT_\lambda$ where for
every marginal state $\lambda\in\Stset(\System_1)$ one chooses a local transformation $\tT_\lambda$ on
$\System_2$ such that $(\tI,\tT_\lambda)\Phi|_1=\lambda$ ($\Phi$ is preparationally faithful with
respect to $\System_2$). Then, one has
\begin{equation}
\forall\lambda\in\Stset(\System_1),\quad
\lambda(a)=(\tI,\tT_\lambda)\Phi(a,e)=
\Phi(a,\tT_\lambda)=\omega_a(\tT_\lambda).
\end{equation}
It follows that $\omega_a=\omega_b$ implies that $\lambda(a)=\lambda(b)$ for all states
$\lambda\in\Stset(\System_1)$, that is $a=b$, whence the homomorphism $a\mapsto\omega_a$ which is
surjective (since $\Phi$ is preparationally faithful) is also injective, {\it i.e.} is bijective, and since
it maps elements of $\Cntset_+(\System_1)$ to elements of $\Stset_+(\System_2)$ and, vice-versa, to
each element of $\Stset_+(\System_2)$ it corresponds an element of $\Cntset_+(\System_1)$ ($\Phi$ is
preparationally faithful), it is a cone-isomorphism. We then have the cone-isomorphism
$\Cntset_+(\System_1)\simeq\Stset_+(\System_2)$. The cone-isomorphism
$\Cntset_+(\System_2)\simeq\Stset_+(\System_1)$ follows by exchanging the two systems.
\item According to point (\ref{i:coneiso12}) one has the cone-isomorphism
$\Cntset_+(\System_1)\simeq\Stset_+(\System_2)\simeq\Stset_+(\System_1)$.
\item Obvious, by definition of preparationally faithful state.
\item According to (\ref{i:coneiso12}) $\omega_a:=\Phi(a,\cdot)$ establishes the cone-isomorphism
  $\Cntset_+(\System)\simeq\Stset_+(\System)$. On the other hand, since the state is both
  preparationally and dynamically faithful for either systems, then for any transformation $\tT$ on
  the first system there exists a unique transformation $\tT'$ on the other system giving the same
  output state (see also the definition of the ``transposed'' transformation with respect to a
  dynamically faithful state in the following). Therefore, since any effect $a$ can be written as
  $a=e\circ\tT_a$ for any $\tT_a\in a$, one has $\omega_a=\Phi(e\circ\tT_a,\cdot)=
  \Phi(e,\cdot\circ\tT_a')=\tT_a'\chi$. The observable-ensemble correspondence and the fact that
  $\chi$ is an internal state are both immediate consequence of the fact that
  $\omega_a:=\Phi(a,\cdot)$ is a cone-isomorphism.  \qed \medskip
\end{enumerate}

\paragraph{\bf Transposed of a transformation} For a symmetric bipartite state $\Phi$ of two identical
systems that is preparationally faithful for one system---hence, according to Theorem \ref{t:faith},
is both dynamically and preparationally faithful with respect to both systems---one can define
operationally the {\bf transposed} $\tT'$ of a transformation $\tT\in\Trnset_\Reals$ through the
identity
\begin{equation}
\Phi(a,b\circ\tT)=\Phi(a\circ\tT',b),
\end{equation}
{\it i.e.} $(\tT',\tI)\Phi=(\tI,\tT)\Phi$, namely, operationally the transposed $\tT'$ of a transformation
$\tT$ is the transformation which will give the same output bipartite state of $\tT$ if operated on the
twin system.  It is easy to verify (using symmetry of $\Phi$) that $\tT''=\tT$ and that
$(\tB\circ\tA)'=\tA'\circ\tB'$. 
\begin{figure}[hbt]
\epsfig{file=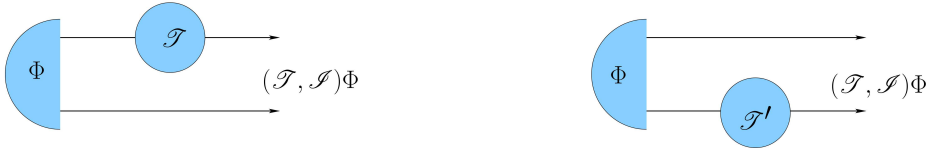,width=12.5cm} 
\caption{Illustration of the notions of transposed of a transformation for a symmetric dynamically
  and preparationally faithful state.}
  \end{figure}

\bigskip

We are now in position to formulate the main Postulate: \Postulate{PFAITH}{Existence of a 
  symmetric preparationally-faithful pure state}{For any couple of identical systems, there exist a
  symmetric (under permutation of the two systems) pure state that is preparationally faithful.}

Theorem \ref{t:faith} guarantees that such a state is both dynamically and preparationally faithful,
and with respect to both systems, as a consequence of symmetry.\footnote{In fact, upon denoting by
  $\tT_\Psi$ the local transformation such that $(\tT,\tI)\Phi=\Psi$, one has
  $(\tI,\tT_{\tS\Psi})\Phi=\Psi$, $\tS$ denoting the transformation swapping the two systems.}
Postulate PFAITH thus guarantees that to any system we can adjoin an ancilla and prepare a pure
state which is dynamically and preparationally faithful with respect to our system. This is
operationally crucial in guaranteeing the preparability of any quantum state for any bipartite
system using only local transformations, and to assure the possibility of experimental calibrability
of tests for any system. Notice that it would be impossible, even in principle, to calibrate
transformations without a dynamically faithful state, since any set of input states
$\{\omega_n\}\in\System'$ that is ``separating'' for transformations $\Trnset(\System')$ is
equivalent to a bipartite state
$\Phi=\sum_n\omega_n\otimes\lambda_n\in\Stset(\System'\odot\System'')$ which is dynamically faithful
for $\System'$, with the states $\{\lambda_n\}$ working just as ``flags'' representing the
``knowledge'' of which state of the set $\{\omega_n\}$ has been prepared.  Notice that in QM every
maximal Schmidt-number entangled state of two identical systems is both preparationally and
dynamically faithful for both systems. In classical mechanics, on the other hand, a state of the
form $\Phi=\sum_l|l\>\< l|\otimes|l\>\< l|$ with $\{|l\>\}$ complete orthogonal set of states (see
footnote \ref{f:refineid}) will be both dynamically and preparationally faithful, however, being not
pure, it would require a (possibly unlimited) sequence of preparations.

On the mathematical side, instead, according to Theorem \ref{t:faith} Postulate PFAITH restricts the
theory to the weakly self-dual scenario ({\it i.e.} with the cone-isomorphism $\Stset_+\simeq\Cntset_+$),
and in finite dimensions one also has the cone-isomorphism
$\Trnset_+(\System)\simeq\Stset_+(\System^{\odot 2})$.  In addition, one also has the following very
useful lemma.

\begin{lemma}\label{l:bip} For finite dimensions Postulate PFAITH implies that the linear space of
  transformations is full, {\it i.e.} $\Trnset_\Field=\Bnd{\Cntset_\Field}$. Moreover, one has
  $\Stset_\Field(\System^{\odot 2})=\Stset_\Field(\System)^{\otimes 2}$ and
  $\Cntset_\Field(\System^{\odot 2})=\Cntset_\Field(\System)^{\otimes 2}$ for
  $\Field=\Reals,\Cmplx$, that is bipartite states and effects are cones spanning the tensor products
  $\Stset_\Field^{\otimes 2}$ and $\Cntset_\Field^{\otimes 2}$, respectively.
\end{lemma}
\Proof In the following we restrict to finite dimensions, with $\Field=\Reals,\Cmplx$ denoting
either the real or the complex fields, respectively.  According to item (2) of Theorem
\ref{t:faith}, for two identical systems the existence of a state that is preparationally faithful
with respect to either one of the two systems implies $\Stset_\Field(\System^{\odot
  2})\simeq\Trnset_\Field(\System)$. Since transformations act linearly over effects one has
$\Trnset_\Field\subseteq\Bnd{\Cntset_\Field}\simeq\Cntset_\Field^{\otimes 2}$, whence
$\Cntset_\Field(\System^{\odot 2})\simeq\Stset_\Field(\System^{\odot
  2})\simeq\Trnset_\Field(\System)\subseteq\Cntset_\Field(\System)^{\otimes 2}$. However, by
local-test embedding one also has $\Cntset_\Field(\System^{\odot
  2})\supseteq\Cntset_\Field(\System)^{\otimes 2}$, whence $\Cntset_\Field(\System^{\odot
  2})=\Cntset_\Field(\System)^{\otimes 2}$, which implies that
$\Trnset_\Field=\Bnd{\Cntset_\Field}$.  Finally, by state-effect duality one also has
$\Stset_\Field(\System^{\odot 2})=\Stset_\Field^{\otimes 2}(\System)$.  \qed \bigskip

The above lemma could have been extended to couples of different systems. However, this
would necessitate with the consideration of more general transformations between different systems 
(see Footnote \ref{f:diffS}).

\bigskip

We conclude that Postulate PFAITH---{\it i.e.} the existence of a symmetric prepara\-tionally-faithful
pure state for bipartite systems---guarantees that we can represent bipartite quantities (states,
effects, transformations) as elements of the tensor product of the single-system spaces. This fact
also implies the following relevant principle

\medskip
\begin{corollary}[Local observability principle] For every composite system there exist
  informationally complete observables made of local informationally complete observables.
\end{corollary}
\Proof A joint observable made of local observables $\AL=\{l_i\}$ on $\System_1$ and $\AM=\{m_j\}$
on $\System_2$ is of the form $\AL\times\AM=\{(l_i,m_j)\}$. Then, by definition, the statement of
the corollary is $\Cntset_\Reals(\System^{\odot 2}) \subseteq\Span_\Reals(\AL\times
\AM)=\Cntset_\Reals^{\otimes 2}(\System)$, which is true according to Lemma \ref{l:bip}.  \qed

\bigskip
Operationally, the Local Observability Principle plays a crucial role, since it reduces enormously
experimental complexity, by guaranteeing that only local (although jointly executed) tests are
sufficient to retrieve a complete information of a composite system, including all correlations
between the components. This principle reconciles holism with reductionism in a non-local theory, in
the sense that we can observe a holistic nature in a reductionistic way, {\it i.e.} locally. 

\bigskip In addition to Lemma \ref{l:bip} and to the local observability principle, Postulate PFAITH has a
long list of remarkable consequences for the probabilistic theory, which are given by the following
theorem.
\begin{theorem}\label{t:pfaith} If PFAITH holds, the following assertions are true
\begin{enumerate}
\item The identity transformation is atomic.
\item\label{i:tansomega} One has $\omega_{a\circ\tA'}=\tA\omega_a$, or equivalently
  $\tA\omega=\Phi(a_\omega\circ\tA',\cdot)$, where $\tA'$ denotes the transposed of $\tA$ with
  respect to $\Phi$. 
\item\label{itransaut} The transpose of a physical automorphism of the set of states is still a physical
  automorphism of the set of states.
\item The marginal state $\chi$ is invariant under the transpose of a channel (deterministic
  transformation) whence, in particular, under a physical automorphism of the set of states.
\item Alice's can perform a perfect EPR-cheating in a perfect concealing bit commitment protocol.
\end{enumerate}
\end{theorem}
\Proof 
\begin{enumerate}
\item According to Theorem \ref{t:faith}-2, the map $\tA\mapsto(\tA,\tI)\Phi$ establishes the
  cone-isomorphism $\Trnset_+\simeq\Stset_+^{\odot 2}$, whence mapping extremal rays of $\Trnset_+$
  to extremal rays of $\Stset_+^{\odot 2}$ and vice-versa it maps the state $\Phi$ itself (which is
  pure) to the identity, which then must be atomic.
\item Immediate definition of the transposition with respect to the dynamically faithful state $\Phi$.
\item Point (\ref{i:tansomega}) establishes that the transposed of a state-automorphism is an effect
  automorphism, which, due to the cone-isomorphism, is again a state-automorphism (see also footnote
  \ref{foot:auto}).
\item For deterministic $\tT$ one has
$\tT'\chi=\Phi(e,\cdot\circ\tT')=\Phi(e\cdot\tT,\cdot)=\Phi(e,\cdot)=\chi$.  The last statement
follows from (\ref{itransaut}) (see also footnote \ref{foot:auto}).
\item{} [For the definition of the protocol, see Ref.\cite{Dariano:2007p27}]. For the protocol to be
  concealing there must exist two ensembles of states $\{\rho_i^\AA\}$ and $\{\rho_i^\AB\}$ that are
  indistinguishable by Bob. These correspond to the two observables $\AA=\{a_i\}$ and $\AB=\{b_i\}$
  with $\rho_i^\AA=\Phi(a_i,\cdot)$ and $\rho_i^\AB=\Phi(b_i,\cdot)$. Instead of sending to Bob a
  state from either one of the two ensembles, Alice can cheat by ``entangling'' her ancilla (system
  $\System_1$) with Bob system in the state $\Phi$, and then measuring either one of the observables
  $\AA=\{a_i\}$ and $\AB=\{b_i\}$.\qed
\end{enumerate}
Notice that atomicity of identity occurs in QM, whereas it is not true in a classical probabilistic
theory (see Footnote \ref {f:refineid}). In classical mechanics one can gain information on the
state without making disturbance thanks to non-atomicity of the identity transformation.  According
to Theorem \ref{t:pfaith}-1 the need of disturbance for gaining information is a consequence of
the purity of the preparationally faithful state, whence disturbance is the price to be payed for
the reduction of the preparation complexity.

\subsection{Scalar product over effects induced by a symmetric faithful state\label{s:scalPhi} }
In this subsection I briefly review the construction in Ref. \cite{darianoVax2007} of a scalar
product over $\Cntset_\Cmplx$ via a symmetric faithful state, along with the corresponding
operational definition of ``transposed'' and ``complex conjugation''---with the composition of the
two giving the adjoint.

According to Theorem \ref{t:faith}-2, for two identical systems in finite dimensions any state that
is preparationally faithful with respect to a system is also dynamically faithful with respect to
the same system. Moreover, according to Postulate PFAITH, there always exists such a state, say
$\Phi$, which is symmetric under permutation of the two systems. The state $\Phi$ is then a
symmetric real form over $\Cntset_\Reals$, whence it provides a non-degenerate scalar product
over $\Cntset_\Reals$ via its Jordan form
\begin{equation}\label{Phimod}
\forall a,b\in\Cntset_\Reals,\quad{}_\Phi( b|a)_\Phi:=|\Phi|(b,a)=\Phi(\varsigma(b),a),
\end{equation}
where $\varsigma$ is the involution $\varsigma=\pi_+-\pi_-$, $\pi_\pm$ denoting the orthogonal
projectors over the positive (negative) eigenspaces of the symmetric form, or, explicitly,
$\varsigma(a):=\sum_j\Phi (a,\tilde f_j)\tilde f_j$ and $\{\tilde f_j\}$ is the canonical Jordan
basis.\footnote{In the diagonalizing orthonormal basis one has $s_j\delta_{ij}=\Phi(\tilde
  f_i,\tilde f_j)=|\lambda_j|^{-1}\Phi(f_i,f_j)$, $s_j=\pm 1$, $\tilde f_j=f_j/\sqrt{|\lambda_j|}$.
} Notice that the Jordan form is representation-dependent---{\em i.e.}  it is defined through the
reference test $\AL=\{l_i\}$---whereas its signature---{\em i.e.} the difference between the numbers
of positive and negative eigenvalues---will be a property of the system $\System$, and will
generally depend on the specific probabilistic theory. For transformations $\tT\in\Trnset_\Reals$ we
define $a\circ\varsigma(\tT):=\varsigma(\varsigma(a)\circ\tT)=: a\circ\tZ\circ\tT\circ\tZ$. For the
identity transformation we have $\varsigma(\tI)=\tZ\circ\tZ=\tI$.  Corresponding to a symmetric
faithful bipartite state $\Phi$ one has the generalized transformation $\tT_\Phi$, given by
\begin{equation}
a\circ\tT_\Phi:=\sum_k\Phi(l_k,a)l_k, 
\end{equation}
for a fixed orthonormal basis $\AL=\{l_j\}$, and in terms of the corresponding symmetric scalar
product $(\cdot,\cdot)_\AL$ introduced in Subsection \ref{s:Cstar}, one has
\begin{equation}
(a,b\circ\tT_\Phi)_\AL=(a\circ\tT_\Phi,b)_\AL=\Phi(a,b).
\end{equation}
Using the dynamical and preparational faithfulness of $\Phi$ we have defined operationally the
transposed $\tT'$ of a transformation $\tT\in\Trnset_\Reals$. Such ``operational'' transposed is
related to the transposed $\tilde\tC$ under the scalar product $(\cdot,\cdot)_\AL$ as
$\tC'=\tT_\Phi\circ\tilde\tC\circ\tT_\Phi^{-1}$. It is easy to check that $\tilde\tZ=\tZ=\tZ'$.

On the complex linear span $\Trnset_\Cmplx$ one can introduce a scalar product as the sesquilinear
extension of the real symmetric scalar product $(\cdot,\cdot)_\Phi$ over $\Cntset_\Reals$ via the
complex conjugation $\eta(\tT)=\tT_R-i\tT_I$, $\tT_{R,I}\in\Trnset_\Reals$, and the adjoint for the
sesquilinear scalar product is then given by
\begin{equation}\label{tomita}
\tT^\dag=\tZ\circ\eta(\tT')\circ\tZ=|\tT_\Phi|\circ\eta(\tilde\tT)\circ|\tT_\Phi|^{-1},
\end{equation}
namely $\tT^\dag=\tZ\circ\tT'\circ\tZ$ on real transformations $\tT\in\Trnset_\Reals$. The Jordan
involution $\varsigma$ thus plays the role of a complex conjugation on $\Trnset_\Reals$, which must
be anti-linearly extended to $\Trnset_\Cmplx$. 

The faithful state $\Phi$ becomes a cyclic and separating vector of a GNS representation by noticing
that $(\tA^{(2)}\Phi)(\eta\varsigma b,a)={}_\Phi( b,a\circ\tA)_\Phi$,\footnote{The action of the
  algebra of generalized transformations on the first system corresponds to the transposed
  representation $(\tA{}^{(1)}\Phi)(\eta\varsigma b,a)=\Phi(\eta\varsigma b\circ\tA,a)=
\Phi(\eta\varsigma b,a\circ\tA')=(\tA'{}^{(2)}\Phi)(\eta\varsigma b,a)$.} and in Eq. (\ref{tomita})
one can recognize the Tomita-Takesaki modular operator of the representation \cite{Haagbook}.

\section{Axiomatic interlude: exploring Postulates FAITHE and PURIFY }
In this section we are exploring two additional postulates of a probabilistic theory: Postulate
FAITHE---the existence of a faithful effect (somehow dual to Postulate PFAITH)---and Postulate
PURIFY---the existence of a purification for every state. As we will see, these new postulates make
the probabilistic theory closer and closer to QM. However, I was still unable to prove (nor to find
counterexamples) that with these two additional postulates the probabilistic theory {\em is} QM.
 
\subsection{FAITHE: a postulate on a faithful effect}
As previously mentioned, Postulate FAITHE is somehow the dual version of Postulate
PFAITH:\footnote{\label{f:PHIPHI} At first sight it seems that the existence of an effect $F$ such
  that $F_{23}\Phi_{12}\Phi_{34}=\alpha\Phi_{14}$ could be derived directly from PFAITH. Indeed,
  according to Lemma \ref{l:bip} for finite dimensions and identical systems we have
  $\Stset_\Field(\System^{\odot 2})=\Stset_\Field(\System)^{\otimes 2}$ and
  $\Cntset_\Field(\System^{\odot 2})=\Cntset_\Field(\System)^{\otimes 2}$ for
  $\Field=\Reals,\Cmplx$.  Moreover, according to Theorem \ref{t:faith}-4 the map
  $a\mapsto\omega_a=\Phi(a,\cdot)$, for $\Phi$ symmetric preparationally faithful achieves the
  cone-isomorphism $\Stset_+\simeq\Cntset_+$, whence for the bipartite system one has
  $\Stset_+(\System^{\odot 2})\simeq\Cntset_+(\System^{\odot 2})$. This leads one to think that it
  should be possible to achieve a preparationally faithful state for $\System^{\odot 4}$ as the
  product $\Phi_{12}\Phi_{34}$. However, this is not necessarily true. In fact, the map
  $\Cntset_\Field(\System)^{\otimes 2}\ni E\mapsto\Omega_E=E_{23}\Phi_{12}\Phi_{34}$ is a linear
  bijection between $\Cntset_\Field(\System)^{\otimes 2}$ and $\Stset_\Field(\System)^{\otimes 2}$
  [since
  $\Span_\Field\{\Phi_{12}(\cdot,a)\Phi_{34}(b,\cdot)|a,b\in\Cntset\}=\Stset_\Field(\System)^{\otimes
    2}= \Stset_\Field(\System^{\odot 2})$], is cone-preserving, it sends separable effects to
  separable states, whence it sends non-separable effects to non-separable states (since it is
  one-to-one). However, it doesn't necessarily achieve the cone-isomorphism $\Stset_+(\System^{\odot
    2})\simeq\Cntset_+(\System^{\odot 2})$, since it is not necessarily true that any bipartite state
  $\Omega$ is the mapped of a bipartite effect $E_\Omega$ (we remember that a cone-isomorphism is a
  bijection that preserves the cone {\em in both directions}). If by chance this would be the
  case---{\it i.e.} $E\mapsto\Omega_E$ is a cone-isomorphism for $\System^{\odot 2}$---then this means that
  there exists an effect $F\in\Cntset(\System^{\odot 2})$---such that $\Omega_F=\alpha\Phi$, with
  $0<\alpha\leq 1$.}

\Postulate{FAITHE}{Existence of a faithful effect}{There exist a bipartite effect
  $F\in\Cntset(\System^{\odot 2})$ achieving the inverse of the isomorphism
  $a\mapsto\omega_a:=\Phi(a,\cdot)$. More precisely
\begin{equation}\label{e:FAITHE}
F_{23}(\omega_a)_2=F_{23}\Phi_{12}(a,\cdot)=\alpha a_3,\quad 0<\alpha\leq 1.
\end{equation} 
} 

Notice that, since $\Phi$ establishes an isomorphism between the cones of states and effects, there
must exist a generalized effect $F\in\Cntset_\Reals^{\otimes 2}$ satisfying Eq. (\ref{e:FAITHE}),
but we are not guaranteed that it is a physical, {\em i.e.} $F\in\Cntset_+(\System^{\odot 2})$.

\bigskip
Let's denote by $\hat F=\alpha^{-1}F$ the rescaled effect in the cone.  Eq. (\ref{e:FAITHE}) can
be rewritten in different notation as follows
\begin{equation}\label{EPHIa}
\hat F(\omega_a,\cdot)=\hat F(\Phi(a,\cdot),\cdot)=a
\end{equation}
\begin{equation}\label{PhiE}
\Phi(a_\omega,\cdot)=\Phi(\hat F(\omega,\cdot),\cdot)=\omega.
\end{equation}
[One needs to be careful with the notation in the multipartite case, {\it e.g.} in Eq. (\ref{PhiE})
$\Phi(\hat F(\omega,\cdot),\cdot)=\omega$ is actually a state, since $\hat F(\omega,\cdot)$ is an
effect, etc.] Both faithful state $\Phi$ and faithful effect $F$ can be used to express the
state-effect pairing, namely
\begin{equation}
\zeta(b)=\Phi(a_\zeta,b)=\hat F(\omega_b,\zeta),\qquad
a_\zeta:=\hat F(\zeta,\cdot),\quad\omega_b:=\Phi(b,\cdot),
\end{equation}
or, substituting
\begin{equation}
\zeta(b)=\Phi(\hat F(\zeta,\cdot),b)=\hat F(\Phi(b,\cdot),\zeta).
\end{equation}

\begin{figure}[h]
\epsfig{file=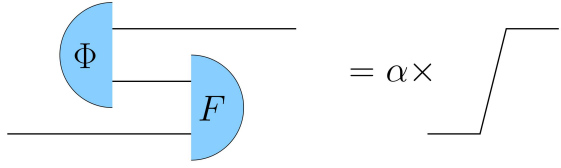,width=8cm}
\caption{Illustration of Postulate FAITHE.\label{f:FAITHE}}
\end{figure}

Eq. (\ref{e:FAITHE}) can also be rewritten as follows
\begin{equation}\label{e:tele}
F_{23}\Phi_{12}=\alpha\,\mathbf{Swap}_{13},
\end{equation}
where $\mathbf{Swap}_{ij}$ denotes the transformation swapping $\System_i$ with $\System_j$.  In
Fig. \ref{f:FAITHE} Postulate FAITHE is illustrated graphically. 

Eq. (\ref{e:tele}) means that using the state $\Phi$ and the effect $F$ one can achieve
probabilistic {\bf teleportation} of states from $\System_2$ to $\System_4$. In fact, one has
\begin{equation}
F_{23}\omega_2\Phi_{34}=F_{23}\Phi_{12}(a_\omega,\cdot)\Phi_{34}=\alpha\Phi_{14}(a_\omega,\cdot)=\alpha\omega_4.
\end{equation}
Using the last identity we can also see that Postulate FAITHE is also equivalent to the identity
\begin{equation}\label{alphaphi}
F_{23}\Phi_{12}\Phi_{34}=\alpha\Phi_{14},
\end{equation}
which by linearity is extended from local effects to all effects, in virtue of $\Cntset^{\odot
  2}=\Cntset^{\otimes 2}$. With equivalent notation we can write $(\Phi,\Phi)(\cdot,F,\cdot)=\alpha\Phi$.

\medskip The effect $F$ is also completely faithful, in the sense that the correspondence
$F_\tA:=F\circ(\tA',\tI)\Longleftrightarrow\tA$ is bijective (in finite dimensions). In fact one has
\begin{equation}
[F\circ(\tA',\tI)]_{23}(\Phi,\Phi)=\alpha(\tA,\tI)\Phi,
\end{equation}
and since $\Phi$ is dynamically faithful (it is symmetric preparationally faithful), the
correspondence $F_\tA:=F\circ(\tA',\tI)\Longleftrightarrow\tA$ is one-to-one and surjective, whence
it is a bijection (in finite dimensions). It is also easy to see that $F\circ(\tA',\tI)]=F\circ(\tI,\tA)$, since
\begin{equation}
\begin{split}
[F\circ(\tI,\tA)]_{23}(\Phi,\Phi)&=F_{23}(\Phi,(\tA,\tI)\Phi)=
F_{23}(\Phi,(\tI,\tA')]\Phi)\\&=\alpha(\tI,\tA')\Phi
=\alpha(\tA,\tI)\Phi=[F\circ(\tA',\tI)]_{23}(\Phi,\Phi),
\end{split}
\end{equation}
whence transposition can be equivalently defined with respect to the faithful effect $F$.  The
bijection $F_\tA:=F\circ(\tI,\tA)\Longleftrightarrow\tA$ is cone-preserving in both directions, since
to every transformation it corresponds an effect, and to each effect $A\in\Cntset(\System^{\odot2
})$ it corresponds a transformation, since
\begin{equation}
A_{23}(\Phi,\Phi)=\Omega_A=(\tT_{\Omega_A},\tI)\Phi=:(\tT_A,\tI)\Phi.
\end{equation}
Therefore, the map $\tA\mapsto F_\tA$ realizes the cone-isomorphism $\Cntset_+(\System^{\odot
  2})\simeq\Trnset_+(\System)$ which is just the composition of the weak-selfduality and of the
isomorphism $\Stset_+(\System^{\odot 2})\\ \simeq\Trnset_+(\System)$ due to PFAITH.
However, as mentioned in footnote \ref{f:PHIPHI}, the map 
\begin{equation}
\Cntset_+(\System^{\odot 2})\ni A\mapsto \Omega_A:=A_{23}(\Phi,\Phi)\in\Stset(\System^{\odot 2}),
\end{equation}
is bijective between $\Stset_\Field(\System^{\odot 2})$ and $\Cntset_\Field(\System^{\odot 2})$,
but it does not realize the cone-isomor\-phism $\Stset_+(\System^{\odot
  2})\simeq\Cntset_+(\System^{\odot 2})$, since it is not surjective over $\Cntset_+(\System^{\odot
  2})$.  Indeed, for $A\in\Cntset(\System^{\odot 2})$ physical effect, one has
$A_{23}(\Phi,\Phi)=(\tT_A,\tI)\Phi$ with $\tT_A\in\Trnset(\System)$ physical transformation.
However, there is no guarantee that, vice-versa, a physical transformation always has a
corresponding physical effect, {\it e.g.} for the identity transformation in Eq.  (\ref{alphaphi}).
It also follows that any bipartite observable $\AA=\{A_l\}$ leads to the {\bf totally depolarizing
  channel} $\tT_{(e,e)}\omega=\chi$, $\forall\omega\in\Stset$.\footnote{Indeed, one has
  $\sum_l(A_l)_{23}\omega_2\Phi_{34}=(e,e)_{23}\omega_2\Phi_{34}=
  \Phi_{12}(a_\omega,e)\Phi_{34}(e,\cdot)=\omega(e)\chi$.}  Using the faithfulness of $F$ it is
possible to achieve probabilistically any transformation on a state $\omega$ by performing a joint
test on the system interacting with an ancilla, {\it i.e.}
$(\omega\Phi)(F_{\tA'},\cdot)=\alpha\tA\omega$ (for Stinespring-like dilations in an operational
context see Ref. \cite{Chiribella:2009unp}).

\bigskip
\paragraph{\bf More about the constant $\alpha$}
Notice that the number $0<\alpha\leq 1$ is the probability of achieving teleportation
$\alpha=(F_{23}\omega_2\Phi_{34})(e)$. It is independent on the state $\omega$, and depends only on
$F$, since it is given by $\alpha\equiv\alpha_F=[F_{23}\Phi_{12}\Phi_{34}](e,e)$. The maximum value
maximized over all bipartite effects
\begin{equation}
\alpha(\System)=\max_{A\in\Cntset(\System^{\odot
    2})}\{(\Phi,\Phi)(e,A,e)\}
\end{equation}
is a property of the system $\System$ only, and depends on the particular probabilistic theory. 

\bigskip

\paragraph{\bf More on the relation between Postulates PFAITH and FAITHE} Postulate PFAITH
guarantees the existence of a symmetric preparationally faithful state for each pair of identical
systems $\System^{\odot 2}$. Now, consider the bipartite system $\System^{\odot 2}\odot
\System^{\odot 2}$, and denote by ${\mathbf\Phi}$ a symmetric preparationally faithful state for
it. The map $A\mapsto\Omega_A:={\mathbf\Phi}(A,\cdot,\cdot)$ $\forall A\in\Cntset(\System^{\odot 2})$
establishes the state--effect cone-isomorphism for $\System^{\odot 2}$, whence there must exists an
effect $A_\Phi$ such that
\begin{equation}
{\mathbf\Phi}(A_\Phi,\cdot,\cdot)=\beta\Phi,\quad 0<\beta\leq 1.
\end{equation}
Suppose now that the faithful state can be chosen in such a way that it maps separable states to
separable effects as follows
\begin{equation}\label{e:factorE}
{\mathbf\Phi}(\cdot,\cdot,(a,b))=\gamma(\omega_a,\omega_b)=\gamma\Phi(\cdot,a)\Phi(\cdot,b),
\;\gamma>0.
\end{equation}
Then one has
\begin{equation}
\gamma (A_\Phi)_{13}(\Phi,\Phi)={\mathbf\Phi}(A_\Phi,\cdot,\cdot)=\beta\Phi,
\end{equation}
namely, according to Eq. (\ref{alphaphi}) one has $\beta^{-1}\gamma A_\Phi\equiv\hat F$, which is
the effect whose existence is postulated by FAITHE. Notice, however, that the factorization Eq.
(\ref{e:factorE}) doesn't need to be satisfied. In other words, the automorphism relating the
cone-isomorphism induced by ${\mathbf\Phi}$ with another cone-isomorphism that preserves local
effects may be unphysical (see also footnote \ref{f:PHIPHI}). One can instead require a stronger
version of postulate PFAITH, postulating the existence of a preparationally {\bf super-faithful}
symmetric state $\Phi$, also achieving a four-partite preparationally symmetric faithful state
${\mathbf\Phi}$ as $(\Phi,\Phi)={\mathbf\Phi}$. A weaker version of such postulate is thoroughly
analyzed in Ref.  \cite{Chiribella:2009unp}, where it is also shown that it leads to
Stinespring-like dilations of deterministic transformations.

\medskip
\paragraph{\bf The case of QM} It is a useful exercise to see how the present framework translates
in the quantum case, and find which additional constraints can arise from a specific probabilistic
theory.  For simplicity we consider a maximally entangled state (with all positive amplitudes in a
fixed basis) as a preparationally symmetric state $\Phi$. The corresponding marginal state is given
by the density matrix $d^{-1}I$, $I$ denoting the identity on the Hilbert space.  For the constant
$\alpha$ one has $\alpha=d^{-2}$, where $d$ is the dimension of the Hilbert space.  A simple
calculation shows that the identity $\omega_a=\tT_a'\chi$ for $\tT_a\in a$ translates
to\footnote{For $\Phi=d^{-1}\sum_{nm}|n\>|n\>\<m|\<m|$ the marginal state is $\chi=d^{-1}I$ and the
  Jordan involution is the complex conjugation with respect to the orthonormal basis $\{|n\>\}$. For
    quantum operation $\tT=\sum_n T_n\cdot T_n^\dag$ with corresponding effect 
$a=\sum_nT_n^\dag T_n$, one has $\tT'\chi=d^{-1}\sum_n\transp{T_n}T_n^*=d^{-1} \sum_n(T_n^\dag
  T_n)^*=\sqrt{\alpha}\varsigma(a)$.}
\begin{equation}\label{e:selfd}
\omega_a=\sqrt{\alpha}\varsigma(a),\qquad\Leftarrow\text{in QM}
\end{equation}
where the involution $\varsigma$ of the Jordan form in Eq. $\Phi$ (\ref{Phimod}) here is also an
automorphism of states/effects, whence identity (\ref{e:selfd}) expresses the self-duality of QM.
Rewriting Eq.  (\ref{e:selfd}) in terms of the faithful effect $F$ (which would be an element of a
Bell measurement), one obtains\footnote{In fact, one has
$\omega_a:=\Phi(a,\cdot)=\sqrt{\alpha}\varsigma(a)$, namely
$\Phi(\varsigma(a),\cdot)=\sqrt{\alpha}a$, {\it i.e.}
$|\Phi|(a,\cdot)=\sqrt{\alpha}a$, and using Eq.  (\ref{EPHIa}) one has
$\sqrt{\alpha}\hat F(\Phi(a,\cdot),\cdot)=|\Phi|(a,\cdot)$, namely the statement.}
\begin{equation}
(\cdot, F)(\Phi,\cdot)=\sqrt{\alpha}|\Phi|,\qquad\Leftarrow\text{in QM}.
\end{equation}
Another feature of QM is that the preparationally faithful symmetric state $\Phi$ is super-faithful,
namely ${\mathbf\Phi}=(\Phi,\Phi)$ is preparationally faithful for $\System^{\odot 4}$.

\subsection{PURIFY: a postulate on purifiability of all states}
In the present section for completeness I briefly explore the consequences of assuming purifiability
for all states, namely: \Postulate{PURIFY}{Purifiability of states}{For every state $\omega$ of
  $\System$ there exist a pure bipartite state $\Omega$ of $\System^{\odot 2}$ having it as marginal
  state, namely
\begin{equation}\label{e:PURIFY}
\forall\omega\in\Stset(\System),\; \exists \Omega\in\Stset(\System^{\odot 2}) \text{ pure, such that
}\Omega(e,\cdot)=\omega.
\end{equation} }
Postulate PURIFY has been analyzed in Ref. \cite{Chiribella:2009unp}, where the following Lemma is
proved
\begin{lemma}\label{l:PURIFY} If Postulate PFAITH holds, then Postulate PURIFY implies the following
  assertions
\begin{enumerate}
\item Even without assuming purity of the preparationally faithful state $\Phi$, the identity
  transformation is atomic, and purity of $\Phi$ can be derived.
\item $\Stset_+\equiv\Erays(\Trnset_+)\chi$, {\it i.e.} each state can be obtained by applying an atomic
  transformations to the marginal state $\chi:=\Phi(e,\cdot)$.
\item $\Cntset_+\equiv e\circ\Erays(\Trnset_+)$, {\it i.e.} each effect can be achieved with an atomic
  transformation.
\end{enumerate}
\end{lemma}

Points (2) and (3) corresponds to the square-root of states and effects in the quantum case.

\section{What is special about Quantum Mechanics as a 
probabilistic theory} 
The mathematical representation of the operational probabilistic framework derived up to now is
completely general for any fair operational framework that allows local tests, test-calibration, and
state preparation. These include not only QM and classical-quantum hybrid, but also other
non-signaling non-local probabilistic theories such as the {\em PR-boxes} theories
\cite{Popescu:1994p1581}. Postulate PFAITH has proved to be remarkably powerful, implying (1) the
local observability principle, (2) the tensor-product structure for the linear spaces of states and
effects, (3) weak self-duality, (4) realization of all states as transformations of the marginal
faithful state $\Phi(e,\cdot)$, (5) locally indistinguishable ensembles of states corresponding to
local observables---{\it i.e.} impossibility of bit commitment---and more. By adding FAITHE one 
even has teleportation!  However, despite all these positive landmarks, it is still unclear if one can derive
QM from these principles only.

\bigskip
What is then special about QM? The peculiarity of QM among probabilistic
operational theories is: \bigskip
\begin{center}
  {\bf Effects not only can be linearly combined, but also they can be composed each other, 
so that complex effect make a C${}^*$-algebra}.
\end{center}
\bigskip Operationally the last assertion is odd, since {\em the notion of effect abhors composition!} Therefore,
the composition of effects ({\it i.e.} the fact that they make a C${}^*$-algebra, {\it i.e.} an operator algebra
over complex Hilbert spaces) must be derived from additional postulates. What I will show
here is: \bigskip
\begin{center}
  {\bf With a single mathematical postulate, and assuming atomicity of evolution, one can derive 
the composition of effects in terms of composition of atomic events}.
\end{center}
\bigskip One thus is left with the problem of translating the remaining mathematical postulate into
an operational one.  Let's now examine the two postulates.

\Postulate{AE}{Atomicity of evolution}{The composition of atomic transformations is atomic.}  This
postulate is so natural that looks obvious.\footnote{Indeed, when joining events $\tA$ and $\tB$
  into the event $\tA\wedge\tB$, the latter is atomic if both $\tA$ and $\tB$ are atomic.} However,
even though for atomic events $\tA$ and $\tB$ the event $\tC=\tB\circ\tA$ is not refinable in the
corresponding cascade-test, there is no guarantee that $\tC$ is not refinable in any other test. We
remember that mathematically atomic events belong to $\Erays(\Trnset_+)$, the extremal rays of the
cone of transformations.

\medskip We now state the mathematical Postulate: \MPostulate{CJ}{Choi-Jamiolkowski isomorphism}{The
  cone of transformations is isomorphic\footnote{For the definition of cone-isomorphisms, see
    Footnote \ref{f:isocones}.} to the cone of positive bilinear forms over complex effects
  \cite{choi75,Jamiolkowski72}, {\it i.e.} $\Trnset_+\simeq\Bndp{\Cntset_\Cmplx}$.}

\begin{figure}[h]
\epsfig{file=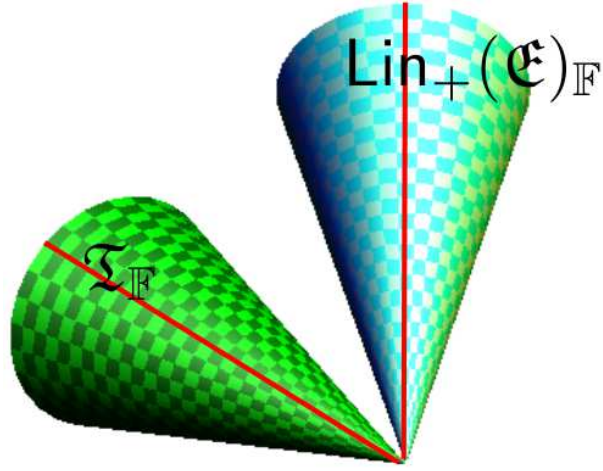,width=8cm}
\caption{The Choi-Jamiolkowski isomorphism between the cone $\Trnset_+$ of physical transformations
  and the cone $\Bndp{\Cntset_\Cmplx}$ of positive matrices over complex effects establishes a
  one-to-one correspondence between extremal ray points of the two cones, identifying effects
  (modulo a phase) with atomic transformations (the lines over the cones represent a pair of
  corresponding rays).}\label{fig:CJ}
\end{figure}

In terms of a sesquilinear scalar product over complex effects, positive bilinear forms can be
regarded as a positive matrices over complex effects, {\it i.e.} elements of the cone
$\Bndp{\Cntset_\Cmplx}$. 

\medskip\par The extremal rays $\Erays(\Bndp{\Cntset_\Cmplx})$ are rank-one positive
operators $|x\>\< x|\in\Erays(\Bndp{\Cntset_\Cmplx})$ with $x\in\Cntset_\Cmplx$, and the map
$\pi:x\mapsto\pi(x):=|x\>\< x|$ is surjective over $\Erays(\Bndp{\Cntset_\Cmplx})$. One has
$\pi(xe^{i\phi})=\pi(x)$, and $\pi^{-1}(|x\>\< x|)=\{e^{i\phi}x\}\subseteq\Cntset_\Cmplx$, {\it i.e.} the
set of complex effects mapped to the same rank-one positive operator is the set of complex effects
that differ only by a multiplicative phase factor.  We will denote by $|x|\in\Cntset_\Cmplx$ a fixed
choice of representative for such an equivalence class,\footnote{An example of choice of representative
  is given by $||x|\>:=\<e_{\iota(x)}|\pi(x)|e_{\iota(x)}\>^{-\frac{1}{2}}\pi(x)|e_{\iota(x)}\>$,
  namely $|x|:=|(x,e_{\iota(x)})|^{-1}(x,e_{\iota(x)})x$, with with $\iota(x)=\min\{i:(x,e_i)\neq
  0\}$, for given fixed basis for $\Cntset_\Cmplx$.\label{f:phi}} introduce the phase corresponding to such
choice as $x=:|x|e^{i\phi(x)}$, and denote by $\Cntset_\Cmplx/\phi$ the set of equivalence classes,
or, equivalently, of their representatives. 
Now, since the representatives $|x|\in\Cntset_\Cmplx/\phi$ are in one-to-one correspondence with the
points on $\Erays(\Bndp{\Cntset_\Cmplx})$, the CJ isomorphism establishes a bijective map between
$\Cntset_\Cmplx/\phi$ and $\Erays(\Trnset_+)$ as follows
\begin{equation}
\tau:\;\Cntset_\Cmplx/\phi\ni|x|\leftrightarrow\tau(|x|)\in\Erays(\Trnset_+).
\end{equation}
\subsection{Building up an associative algebra structure for complex effects\label{s:associative}} 
Assuming Postulate AE, we can introduce an associative composition between the effects in
$\Cntset_\Cmplx/\phi$ via the bijection $\tau$
\begin{equation}\label{|a||b|}
|a||b|:=\tau^{-1}(\tau(|a|)\circ\tau(|b|)).
\end{equation}
Notice that, by definition, $|a||b|$ is a representative of an equivalence class in
$\Cntset_\Cmplx$, whence $|\,(|a||b|)\,|=|a||b|$. The above composition extends to all elements
of $\Cntset_\Cmplx$ by taking
\begin{equation}\label{ab}
ab:=|a||b|e^{i\phi(a)}e^{i\phi(b)},
\end{equation}
and since $|\,(|a||b|)\,|=|a||b|$, one has $|ab|=|a||b|$, and $\phi(ab)=\phi(a)+\phi(b)$. It follows that
the extension is itself associative, since
\begin{equation}
\begin{split}
(ab)c=&|ab||c|e^{i\phi(ab)+i\phi(c)}=|a||b||c|e^{i\phi(a)+i\phi(b)+i\phi(c)}\\
=&|a||bc|e^{i\phi(a)+i\phi(bc)}=a(bc).
\end{split}
\end{equation}
The composition is also distributive with respect to the sum, since it follows the same rules of
complex numbers. We will denote by $\iota$ the identity in $\Cntset_\Cmplx/\phi$ when it exists,
which also works as an identity for multiplication of effects as in Eq. (\ref{ab}). Notice that since
the identity transformation $\tI$ is atomic, one has $\iota:=\tau^{-1}(\tI)\in\Cntset_\Cmplx/\phi$
according to Eq. (\ref{|a||b|}). 

\subsection{Building up a C${}^*$-algebra structure over complex effects}
We want now to introduce a notion of adjoint for effects. We will do this in two steps: (a) we
introduce an antilinear involution on the linear space $\Cntset_\Cmplx$; (b) we extend the
associative product (\ref{ab}) under such antilinear involution. 
\begin{itemize}
\item[(a)] First we notice that the complex space $\Cntset_\Cmplx$ has been constructed as
  $\Cntset_\Cmplx=\Cntset_\Reals\oplus i\Cntset_\Reals$ starting from real combinations of physical
  effects $\Cntset_\Reals=\Span_\Reals(\Cntset_+)$, {\it i.e.} one has the unique Cartesian decomposition
  $x=x_R+ix_I$ of $x\in\Cntset_\Cmplx$ in terms of $x_R,x_I\in\Cntset_\Reals$. We can then define
  the antilinear {\em dagger} involution $\dag$ on $\Cntset_\Cmplx$ by taking $x^\dag=x$ $\forall
  x\in\Cntset_\Reals$ and $x^\dag:=x_R-ix_I$ $\forall x\in\Cntset_\Cmplx$. Notice that
  $\Cntset_\Cmplx$ is closed under such involution. Taking the involution of the defining identity
  $x=:|x|e^{i\phi(x)}$ one has $|x^\dag|=|x|^\dag e^{-i\phi(x^\dag)-i\phi(x)}$ which is consistently
  satisfied by choosing $|x^\dag|=|x|^\dag$ and $\phi(x^\dag)=-\phi(x)$ $\forall x\in\Cntset_\Cmplx$
  (these identities are satisfied {\it e.g.} for the choice of representative in Footnote \ref{f:phi}).
\item[(b)] The multiplications $a^\dag b$ and $ab^\dag$ are defined via the scalar product over
  $\Cntset_\Cmplx$ as follows\footnote{The right and left multiplications are just special elements
    of the algebra $\Bnd{\Cntset_\Cmplx}$, whence their adjoints are definable via the scalar
    product as usual.}
\begin{equation}\label{defadj}
\forall c\in\Cntset_\Cmplx:\qquad(c,a^\dag b):=(ac,b),\qquad(c,ab^\dag):=(cb,a). 
\end{equation}
This is possible since the scalar product over $\Cntset_\Cmplx$ is supposed to be non-degenerate.
It is then easy to verify that one has the identities $(ab)^\dag=b^\dag a^\dag$ and $\iota^\dag=\iota$. 
\end{itemize}
In this way $\Cntset_\Cmplx$ is closed under complex linear combinations, adjoint, and associative
composition, and possibly contains the identity element $\iota$, that is it is an associative complex
algebra with adjoint, closed with respect to the adjoint. The scalar product on $\Cntset_\Cmplx$ in
conjunction with the identity, leads to a strictly positive linear form over $\Cntset_\Cmplx$,
defined as $\Phi=(\iota,\cdot)$, and one has $\Phi(a^\dag b)=(\iota,a^\dag b)=(a,b)$.\footnote{The
  form is strictly positive since $\Phi(a^\dag a)=(a,a)\geq 0$, with the equal sign only if $a=0$,
  since the scalar product is non-degenerate.}  Such form is also a {\em trace}, {\it i.e.} it satisfies
the identity $\Phi(ba)=\Phi(ab)$, which can be easily verified using definitions
(\ref{defadj}).\footnote{One has $\Phi(ab)=(\iota,ab)=(\iota,a(b^\dag)^\dag)=(b^\dag,a)$ and
  $\Phi(ba)=(\iota,ba)=(\iota,(b^\dag)^\dag a)=(b^\dag,a)$.}  The complex linear space of the
algebra closed with respect to the norm induced by the scalar product makes it a Hilbert space, and
the action of the algebra over itself regarded as a Hilbert space makes it an operator
algebra.\footnote{This construction is a special case of the Gelfand-Naimark-Segal (GNS)
  construction\cite{GelfandNeumark}, in which the form $\Phi$ is a trace. In the standard GNS
  construction the form $\Phi$ maybe degenerate, {\it i.e.} one can have $\Phi(a^\dag a)=0$ for some $a\neq
  0$, and the vectors of the representation are built up as equivalence classes modulo vectors
  having $\Phi(a^\dag a)=0$.}  It is a standard result of the theory of operator algebras that the
closure of $\Cntset_\Cmplx$ under the operator norm (which is guaranteed in finite dimensions) is a
C${}^*$-algebra. We have therefore built a C${}^*$-algebra structure over the complex linear space
of effects $\Cntset_\Cmplx$.  This is the {\em cyclic representation} \cite{Haagbook} given by
\begin{equation}
\Phi(a)=\<\iota|\pi_\Phi(a)|\iota\>,
\end{equation}
$\pi_\Phi$ denoting the algebra representation corresponding to $\Phi$.\footnote{This means that
  $\pi_\Phi(a)\pi_\Phi(b)=\pi_\Phi(ab)$ and $\pi_\Phi(a^\dag)=\pi_\Phi(a)^\dag$.} In our case one
has $\pi_\Phi(a)|\iota\>=|a\>$, along with the {\em trace} property
$\<\iota|\pi_\Phi(a)\pi_\Phi(b)|\iota\>=\<\iota|\pi_\Phi(b)\pi_\Phi(a)|\iota\>$. The latter can be
actually realized as a trace as $\Phi(a^\dag b)=\Tr[O(a)^\dag O(b)]$, via a faithful representation $O:
a\mapsto O(a)\in\Bnd{H}$ of the algebra $\Cntset_\Cmplx$ as a subalgebra of $\Bnd{H}$ of operators
over a Hilbert space $\sH$ with dimension $\dim(\sH)^2\geq\dim(\Cntset_\Cmplx)$. In this way,
one has $\pi_\Phi(a)=(O(a)\otimes I)$ with the cyclic vector represented as
$|\iota\>=\sum_n|n\>\otimes|n\>$, $\{|n\>\}$ being any orthonormal basis for $\sH$.
\subsection{Recovering the action of transformations over effects}
In order to complete the mathematical representation of the probabilistic theory, we now need to
define the action of the elements of $\Trnset_\Cmplx$ over $\Cntset_\Cmplx$, and to select the cone
of physical transformations $\Trnset_+$. We will show that $\Trnset_+$ is given by the completely
positive linear maps on $\Cntset_\Cmplx$, namely the linear maps of the Kraus form, {\it i.e.} the atomic
transformations act on $x\in\Cntset_\Cmplx$ as $x\circ\tau(|a|)=|a|^\dag x |a|\equiv a^\dag x a$.

\begin{figure}[h]
\epsfig{file=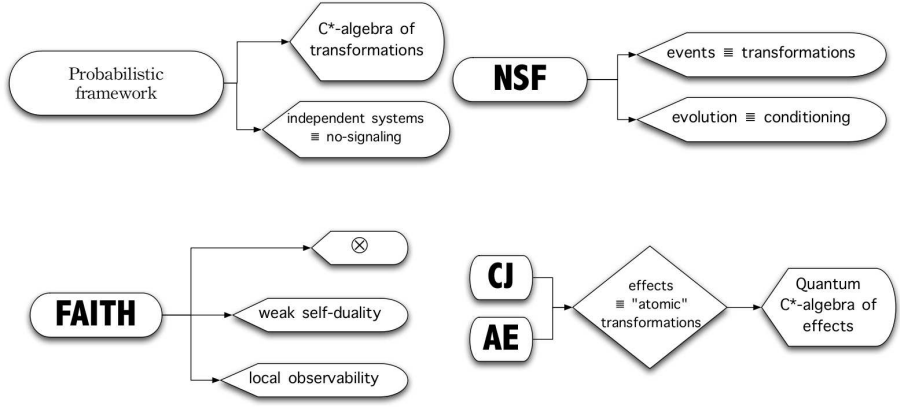,width=12cm}
\caption{Operational axiomatic framework for quantum Mechanics: summary of the relevant logical
  implications.}\label{fig:AX}
\end{figure}

First, notice that the full span $\Bnd{\Cntset_\Cmplx}$ is recovered from
$\Erays(\Bndp{\Cntset_\Cmplx})$ via the polarization identity
\begin{equation}\label{dyads}
|a\>\< b|=\frac{1}{4}\sum_{k=0}^3i^k|(a+i^kb)\>\<(a+i^kb)|.
\end{equation}
Correspondingly, we introduce the generalized transformations
\begin{equation}
\tau(b,a):=\frac{1}{4}\sum_{k=0}^3i^k\tau(|a+i^kb|)\in\Trnset_\Cmplx.
\end{equation}
The map 
\begin{equation}
|a\>\< b|\mapsto\chi(|a\>\< b|):=b^\dag\cdot a
\end{equation}
is a CJ isomorphism: it represents a bijective map between the cones $\Bndp{\Cntset_\Cmplx}$
and $\Trnset_+$ which can be extended to a cone-preserving linear bijection between
$\Bnd{\Cntset_\Cmplx}$ and $\Trnset_\Cmplx\equiv\Bnd{\Cntset_\Cmplx}$.\footnote{This can be directly
  checked using the operator algebra representation built over $\Cntset_\Cmplx$, whereas the
  isomorphism corresponds to the map $O(b^\dag xa)=\chi(|a\>\<b|)(x)=\Tr_1[(O(x)\otimes I)|a\>\<
  b|]$, and, reversely, $|a\>\<b|=\chi^{-1}(\tau(b,a))=(\tau(b,a)\otimes\tI)(|\iota\>\<\iota|)$.} As
a consequence of Eq. (\ref{|a||b|}), the CJ isomorphism $\tau:|a|\mapsto\tau(|a|)$ will differ from
the isomorphism $\chi$ by an automorphism $\map{U}$ of the C${}^*$-algebra of effects, that is one
has $x\circ\tau(|a|)=\map{U}(a^\dag) x\map{U}(a)$, with $\map{U}(a)=u^\dag au$ with $uu^\dag=u^\dag
u=\iota$.  It follows that the probabilistic equivalence classes are given by
$[\tau(|a|)]_\eff=e\circ\tau(|a|)=u^\dag a^\dag a u$. Notice that
$[\tau(\iota)]_\eff=u^\dag\iota^\dag \iota u=\iota$, that is $\iota$ coincides with the deterministic
effect $\iota=e$.  Complex effects are thus recovered from atomic transformations via the identity
$e\circ\tau(e,a)=u^\dag au$. In Fig \ref{fig:AX} a flow-diagram is reported summarizing the relevant
logical implications of the present operational axiomatic framework for QM.

\subsection{Reconstructing Quantum Mechanics from the probabilistic theory}
It is now possible to reconstruct from the probability tables of the systems the full
C${}^*$-algebra of complex effects $\Cntset_\Cmplx$ as an operator algebra
$\Cntset_\Cmplx\subseteq\oplus_i\Bnd{H_i}$. Here is the recipe:
\begin{enumerate}
\item Look for all sub-cones $(\Cntset_+)_i$ invariant under $\Trnset_+$. 
\item[] Then, for each $i$: 
\item introduce a complex Hilbert space $\sH_i$ such that $(\Cntset_\Cmplx)_i\subseteq
\Bnd{H_i}$, {\it i.e.} with $\dim(\sH_i)=\left\lceil\sqrt{\dim[(\Cntset_\Cmplx)_i]}\right\rceil$, $\lceil x\rceil$ 
  the smallest integer greater than $x$;
\item represent $e$ as the identity over $\oplus_i\sH_i$;
\item build $(\Trnset_\Cmplx)_i\subseteq\Bnd{\Bnd{\sH_i}}$;
\item look for atomic transformations $\Erays(\Trnset_+)_i$;
\item\label{I1} for a given atomic transformation $\tA\in\Erays(\Trnset_+)_i$ take an operator
  $A\in\Bnd{H_i}$ to represent $\tA$ as $A^\dag\cdot A\in\Bnd{\Bnd{\sH_i}}$;
\item\label{I2} represent $[\tA]_\eff$ as $A^\dag A$;
\item\label{I3} repeat steps \ref{I1} and \ref{I2} for another transformation $\tB$;
\item\label{I4} compose $\tC=\tB\circ\tA$ and represent $\tC$ as $C^\dag\cdot C$, with $C=AB$;
\item repeat steps \ref{I3} and \ref{I4} up to build the whole algebra of effects, and the
  corresponding representation of the algebra of transformations;
\item construct states as density operators using the Gleason-like theorem \cite{Gleason57} for
  effects \cite{Busch:2003p1828,Renes00}. 
\end{enumerate}

\begin{center}\begin{table}\begin{tabularx}{\linewidth}{|c|X|X|}
      \hline
      Symbol(s) & Meaning & Related quantities\\ \hline \hline 
$\System_1\odot\System_2$ & Bipartite system obtained by composting $\System_1$ with $\System_2$ & 
\\ \hline $\System=\{\AA,\AB,\AC,\ldots\}$ & System & \\ \hline 
      $\AA,\AB,\AC,\ldots$ &Tests &$\AA=\{\tA_j\}$ Test$:=$ set of possible events\\ \hline
      $\tA,\tB,\tC,\ldots$ & Events$\equiv$transformations & \\ \hline $\omega$ &states, $\Stset$
      Convex set of states
      &$\omega(\tA)$: probability that event $\tA$ occurs in state $\omega$\\
      \hline $\Trnset$ & Convex monoid of transformations/events & $\Trnset_\Reals$,
      $\Trnset_\Cmplx$: linear spans of $\Trnset$, $\Trnset_+$: convex cone\\ \hline $[\tA]_\eff$
      &Effect containing event $\tA$ & \\ \hline $a,b,c,\ldots$ &Effects & $e$: deterministic effect
      \\ \hline $\Cntset$ & Convex set of effects &$\Cntset_\Reals$, $\Cntset_\Cmplx$: linear spans
      of $\Cntset$, $\Cntset_+$: convex cone\\ \hline $\AL=\{l_j\}$ & observable
      &$\sum_{l_i\in\AL}l_i=e$ \\ \hline $\Trnset_\Cmplx$ & C${}^*$-algebra of
      transformations/events &\\ \hline $a\circ\tT$ & Operation of transformation $\tT$ over effect
      $a$ & \\ \hline $\omega_\tA$ &Conditioned states
      &$\omega_\tA:=\omega(\cdot\circ\tA)/\omega(\tA),\qquad$ $\tA\omega=\omega(\cdot\circ\tA)$ \\
      \hline $\Bndp{\Cntset_\Cmplx}$ & Cone of linear maps corresponding to positive bilinear forms
      over $\Cntset_\Cmplx$& $\Bndp{\Cntset_\Cmplx}=$\hfill\break
      $\left\{\tT\in\Trnset_\Cmplx:(a,a\circ\tT)\geq 0,\qquad\right.$ $\left.\forall a\in\Cntset_\Cmplx\right\}$ \\
      \hline
\end{tabularx}
\medskip
\caption{Summary of notation\label{t:notation}}
\end{table}
\end{center}
\section{Conclusions}
Theoretical physics should be, in essence, a mathematical ``representation'' of reality. By
``representation'' we mean to describe one thing by means of another, to connect the object that we
want to understand---the {\em thing-in-itself}---with an object that we already know well---the {\em
  standard}. In theoretical physics we lay down morphisms from structures of reality to
corresponding mathematical structures: groups, algebras, vector spaces, {\it etc.}, each mathematical
structure capturing a different side of reality.

QM  somehow goes differently. We have a beautiful simple mathematical structure---Hilbert
spaces and operator algebras---with unprecedented predictive power in the entire physical domain.
However, we don't have morphisms from the operational structure of reality into a mathematical
structure. In this sense we can say that QM is not yet truly a ``representation'' of reality.
The large part of the formal structure of QM is a set of formal tools for describing the process of
gathering information in any experiment, independently of the particular physics involved. It is
mainly a kind of information theory, a theory about our knowledge of physical entities rather than
about the entities themselves. If we strip off such informational part from the theory, what would
be left should be the true general principle from which QM should be derived.

In the present work I have analyzed the possibility of deriving QM as the mathematical
representation of a fair operational framework made of a set of rules which allows one to make
predictions on future events on the basis of suitable tests. The two Postulates NSF and PFAITH need
to be satisfied by an operational framework that is fair, the former in order to be able to make
predictions based on present tests, the latter to allow calibrability of any test and preparability
of any state.  We have seen that all theories satisfying NSF admit a C${}^*$-algebra representation
of events as linear transformations of complex effects. Based on a very general notion of dynamical
independence, all such theories are {\em non-signaling}.  The C${}^*$-algebra representation of
events is just the informational part of the theory. We have then added Postulate PFAITH.  Postulate
PFAITH has been proved to be remarkably powerful, implying the local observability principle, the
tensor-product structure for the linear spaces of states and effects, weak self-duality, and a list
of features such as realization of all states as transformations of the marginal faithful state
$\Phi(e,\cdot)$, locally indistinguishable ensembles of states corresponding to local
observables---{\it i.e.} impossibility of bit commitment, and more. We have then explored a postulate dual
to PFAITH, Postulate FAITHE for effects, thus deriving additional quantum features, such as
teleportation. We feel that we are really close to QM: maybe we are already there and we only need
to prove it!  All the consequences of these postulates need to be explored further. I have also
reported some consequences of a postulate about the purifiability of all states.  In any case, we
have seen that whichever is the missing postulate, it must establish a one-to-one correspondence
between complex effects and atomic transformations, which, assuming atomicity of evolution
(Postulate AE) will make also effects a C${}^*$-algebra. This is what is special about QM (and all
hybrid quantum-classical theories), and will exclude the other non-signaling probabilistic theories
of the kind of the PR-boxes.\footnote{The PR-boxes in principle satisfy NSF, and can admit a
  dynamical faithful state, {\it e.g.} the boxes of Ref.  \cite{Short:2006p1859} (private discussion with
  Tony Short).} We have seen that the correspondence between effects and atomic transformations is
established by the Choi-Jamiolkowski isomorphism, which is hoped to be not too far from an
operational principle.
\section*{Acknowledgments}
The present research program started in August 2003 in Evanston IL at Northwestern University, where
it continued during many following visits, far from the distractions of normal life, thanks to the
generous hospitality of Horace Yuen. During these years the work has benefited of feedback and
encouragement from many authors and colleagues at various conferences, where preliminary results were
presented, most frequently at V\H{a}xj\H{o} and Perimeter, but also at meetings in Losini, Zurich,
Tokyo, Sendai, Cambridge UK, Obergurgl, and Leiden. In particular the present manuscript was
initiated in Cambridge UK thanks to the hospitality of Robert Spekkens in occasion of the workshop
{\em Operational probabilistic theories as foils to quantum theory}, July 2-13, 2007, and was
finished in Nagoya thanks to the hospitality of Masanao Ozawa in March 2009. I acknowledge useful
discussions with numerous interested fellows, especially with Dirk Schlingeman, Tony Short, Ray
Streater, Chris Fuchs, Marcus Appleby, Karl Svozil, Gregg Jaeger, Lucien Hardy, and valuable
encouragement at the very beginning from Maria Luisa dalla Chiara, Enrico Beltrametti, Guido
Bacciagaluppi and Jos Uffink.  I'm most grateful to my talented disciples Giulio Chiribella and
Paolo Perinotti for pointing out errors in the previous versions of the manuscript, and for numerous
suitable advices.  The work has also greatly gained from a thorough discussion with Alexander Wilce,
Howard Barnum, and Reinhard Werner held in Obergurgl during QICS Workshop on {\em Foundational
  Structures for Quantum Information and Computation} September 14-20, 2008.  Finally, I'm indebted
with Masanao Ozawa and Reinhard Werner for consistently supporting this research and following its
progress, with deep analysis and very useful suggestions.

\end{document}